\documentclass[useAMS,usenatbib]{mn2e}

\usepackage{epsfig}
\usepackage{myaasmacros}
\usepackage{amsmath}

\title[The evolution of the intrinsic scatter of massive black holes]
{On the evolution of the intrinsic scatter in black hole versus galaxy mass relations}
\author[Hirschmann et al.]{Michaela Hirschmann$^{1}$\thanks{E-mail:
mhirsch@usm.lmu.de}, Sadegh Khochfar$^{2}$, Andreas Burkert$^{1,2}$, Thorsten Naab$^{1,3}$,
\newauthor Shy Genel$^{2}$, Rachel S. Somerville$^{4,5}$\\
$^{1}$Universit\"ats-Sternwarte M\"unchen, Scheinerstr.1, D-81679 M\"unchen, Germany\\
$^{2}$Max-Planck-Institut f\"ur extraterrestrische Physik, Giessenbachstrasse, D-85748 Garching, Germany\\
$^{3}$Max-Planck-Institut f\"ur Astrophysik, Karl-Schwarzschild-Strasse 1, D-85741 Garching, Germany\\
$^{4}$Space Telescope Science Insitute, 3700 San Martin Dr., Baltimore, MD 21218, USA\\
$^{5}$Department of Physics and Astronomy, Johns Hopkins University, Baltimore, MD 21218, USA}

\begin{document}

\date{Accepted ???. Received ??? in original form ???}

\pagerange{\pageref{firstpage}--\pageref{lastpage}} \pubyear{2002}

\maketitle

\label{firstpage}

\begin{abstract}
We present results on the evolution of the intrinsic scatter of black hole masses considering different implementations of a model 
in which black holes only grow via mergers. We demonstrate how merger driven growth affects the correlations between black hole 
mass and host bulge mass. The simple case of an initially log-normal distributed scatter in black 
hole and bulge masses combined with random merging within the galaxy population results in a decreasing scatter with merging 
generation/number as predicted by the Central-limit theorem. In general we find that the decrease in scatter $\sigma$ is well  
approximated by $\sigma_{\mathrm{merg}}(m) \sim \sigma_{\mathrm{ini}} \times (m+1)^{-a/2}$ with $a = 0.42$ for a range of mean number of 
mergers $m < 50$. For a large mean number of mergers ($m > 100$) we find a convergence to $a = 0.61$. This is valid for a wide 
range of different initial distributions, refill-scenarios or merger mass-ratios.  Growth scenarios based on halo merger trees of 
a $(100\ \mathrm{ Mpc})^3$ dark matter $\Lambda$CDM-simulation show a similar behaviour with a 
scatter decrease of $a = 0.30$ with typical number of mergers $m < 50$ consistent 
with random merging (best matching model: $a=0.34$). Assuming a present day scatter of $0.3$ dex in black hole mass 
and a mean number of mergers not exceeding $m = 50$ our results imply a scatter of $0.6$ 
dex at $z = 3$ and thus a possible scenario in which overmassive (and undermassive) black holes at high redshift are a consequence of a larger intrinsic scatter in black hole mass.
A simple toy model connecting the growth of black holes to the growth of $\Lambda$CDM dark matter halos via mergers, neglecting any contribution from accretion, yields a consistent $M_{\bullet}-M_{Bulge}$ relation at $z=0$ - if we assume the correct initial relation.

\end{abstract}


\section{Introduction}

There is growing observational evidence that most if not all bulges and elliptical galaxies host a 
supermassive black hole (SMBH) at the present time (\citealp{Magorrian}, \citealp{Genzel}). Furthermore there 
exists a strong correlation between black hole masses and properties of the host galaxy, e.g. the host bulge 
mass and the bulge velocity dispersion (\citealp{Haering}, \citealp{FerrMerr}, \citealp{Gebhardt00}, 
\citealp{Tremaine02}) and possibly the host halo (\citealp{Ferr02}) in nearby galaxies. 
\citet{Haering} find the relation between black hole mass $M_{\bullet}$ and bulge mass $M_{\mathrm{Bulge}}$ to be:
\begin{equation}
\log(M_{\bullet}/M_{\odot}) = 8.20 + 1.12 \times \log(M_{\mathrm{Bulge}}/10^{11} M_{\odot}).
\end{equation}
The correlation between black hole mass $M_{\bullet}$ and velocity dispersion $\sigma_*$ can be written as
\begin{equation} \label{BHsigmarel}
\log(M_{\bullet}/M_{\odot}) = a  \log (\sigma_*/200 \mathrm{km s^{-1}}) + b,
\end{equation}
where a is the slope and b is the zero point. In the literature the values for the slope vary between $a=3.68$ and $a = 4.86$. 
For example \citet{FerrMerr} give values for the zero point $b=-2.9$ and the slope $a = 4.80$. 
\citet{Gebhardt00} claim a smaller slope of $a=3.75$, \citet{Tremaine02}  find $a=4.02$ and more recently 
\citet{Ferrarese05} estimated a slope of $4.86$ and \citet{Graham08} one of $3.68$ for barless galaxies. 
Concerning the intrinsic scatter $\sigma$ in black-hole mass, $\log M_{\bullet}$, most studies agree that 
the scatter is not larger than $0.3\ \mathrm{ dex}$ \citep{Gebhardt00, Tremaine02, Novak06}. Note, that with 
$\sigma_*$ we describe the central velocity dispersion of a galaxy, whereas we characterize the intrinsic 
scatter as $\sigma$.  
In contrast to \citet{Gebhardt00} and \citet{Tremaine02}, a recent study by \citet{Gueltekin09} obtains 
for Eq. \ref{BHsigmarel} a slope $a = 4.24$ and an intrinsic scatter in black hole mass of $\sigma = 0.44\ \mathrm{dex}$ for a  
sample of 49 $M_{\bullet}$-measurements. For a subsample of early-type galaxies they find a smaller 
slope as well as a smaller scatter ($a = 3.96, \sigma = 0.31\ \mathrm{dex}$) than for the full sample.  For non-elliptical galaxies 
the intrinsic scatter is larger with $\sigma = 0.53\ \mathrm{dex}$. In the following the scatter $\sigma$ will always be given in $\mathrm{dex}$.
The existence of these tight correlations strongly suggest a co-evolution of the black hole 
and the bulge component of the host galaxy. However, the origin of these relations is uncertain 
and a subject of current research (e.g. \citealp{Volonteri09, Pengrandom, Burkert01, Springel05, Johansson09}). Apparently, the origin of these relations can be 
connected to the gas dynamics in major galaxy mergers (\citealp{Mihos96}, \citealp{Naab06a}, \citealp{Robertson06a}, 
\citealp{Hopkins08679}). In this scenario, the black holes grow significantly in gas rich mergers of disk galaxies 
and the remnants appear on the observed scaling relations. Subsequent, possibly gas poor, 
major and minor merging (\citealp{Khochfar03}, \citealp{Naab06}, \citealp{Khochfar09}, 
\citealp{Naab09}) conserves the relation (\citealp{Sesana04}, \citealp{Robertson06a}, 
\citealp{Pengrandom}, \citealp{Hopkins07b}, \citealp{Hopkins07a}, \citealp{Springel05}, \citealp{Johansson09}).
\begin{figure}
\begin{center}
  \epsfig{file=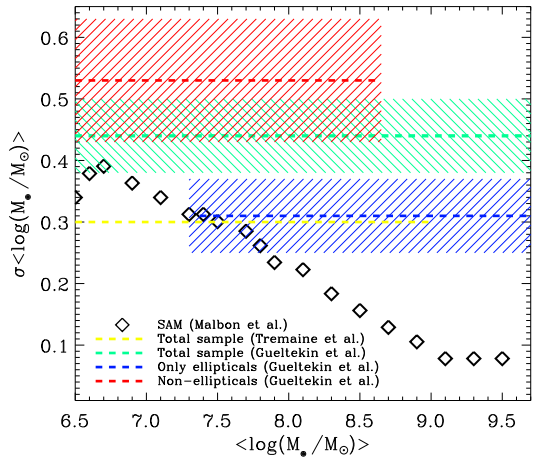, width=0.45\textwidth}
  \caption{Comparison of the intrinsic scatter from observations (\citealp{Tremaine02}, \citealp{Gueltekin09}) 
and theoretical predictions (\citealp{Malbon}). From the 10-90 percentile spread in the model we have 
calculated the corresponding theoretical scatter. The shaded areas illustrate the error on the 
observationally estimated intrinsic scatter.}
  {\label{Comparison}}
\end{center}
\end{figure}

Observationally the black hole mass relations are well constrained only in the nearby universe 
and it is unclear if and how they evolve with cosmic time. Several authors have found evidence that 
galaxies at higher redshift have a higher black hole to bulge mass ratio $M_{\bullet}/M_{\mathrm{Bulge}}$ 
than ellipticals today (\citealp{McLure06, Treu07, Woo08, Walter, Schramm, Pengobs, Greene09, Natarajan, Salviander07, Shields06}). For a sample of Seyfert galaxies at moderate redshifts $z < 0.1$ 
the black holes are more massive by $\Delta \log M_{\bullet} \sim 0.5\ \mathrm{dex}$  compared to the local black 
hole-bulge mass relation (\citealp{Treu07}, \citealp{Woo08}). \citet{Salviander07} find at redshift $z \approx 1$ an evolution of the $M_{\bullet}$-$\sigma_*$-relation by $0.2\ \mathrm{dex}$ in black hole mass. At higher redshifts of $z \sim 2$ \citet{McLure06} 
observe black holes $8$ times more massive than expected and \citet{Pengobs} show that for $z \ge 2$ the 
$M_{\bullet}/M_{\mathrm{Bulge}}$-ratio is $3-6$ times larger than today. This has been confirmed by \citet{Greene09} based on a lensed quasar sample. \citet{Schramm} find evidence 
for an excess in $M_{\bullet}/M_{\mathrm{Bulge}}$ at $z \sim 3$ of a factor of $\sim 10$. At redshifts $4 < z < 6$ \citet{Shields06} obtain for black hole masses in the range of $8 < \log(M_{\bullet}/M_{\odot}) < 10$ a deviation from the present-day $M_{\bullet}$-$M_{\mathrm{bulge}}$-relation of $\Delta \log(M_{\bullet}) \sim 2\ \mathrm{dex}$.
\citet{Walter} report an even higher redshift object, a quasar at $z=6$, whose black hole is about $20$ 
times more massive than expected. Considering a present-day scatter according to 
\citet{Tremaine02} ($\sigma = 0.3$) or according to \citet{Gueltekin09} ($\sigma = 0.31$ for ellipticals) 
all observed black holes at $z \ge 2$ are outside the $2-\sigma$ range of the present-day scatter.
Furthermore, recent observations (\citealp{Alexander08, Shapiro09}) show also the existence of under-massive black holes at high redshifts. \citet{Alexander08} find black hole masses, which are $3$ times smaller than those found in comparable massive galaxies 
in the local Universe. The results in \citet{Shapiro09} show that black hole masses at $z=2$ are an order of magnitude lower than 
those predicted by local scaling relations.  

The most obvious explanation for the over-massive black holes at high redshifts are possible selection 
effects. It is more likely to detect the most luminous and most massive black holes at high redshift 
than less luminous ones. However, the probability for finding a massive black hole in 
the mass range $10^9 - 10^{10} M_{\odot}$ in the observed volume at $z = 3$ (e.g. \citealp{Schramm}) is extremely 
low as estimates of the local SMBH mass function from SDSS (\citealp{Benson07}) would predict no high mass black holes 
to be found in a similar volume even assuming no evolution in the SMBH mass function. Cosmic variance is very unlikely to be an explanation for the observed, massive black holes. \citet{Lauer07} point out that there is an additional bias which is due to different selection effects for high-redshift (e.g. black holes in high-z galaxies selected by nuclear activity) and local samples (e.g. black holes in local galaxies selected by luminosity or velocity dispersion). They deduce that because of this bias $M_{\bullet}$ will typically appear to be too large in a distant sample for a given luminosity or velocity dispersion. 
Some authors (e.g. \citealp{Croton06}) explain observed, over-massive black holes at high redshift with 
a shifted relation towards higher black hole masses for a given bulge mass. He uses the Millennium $\Lambda$CDM-simulation 
(\citealp{Springel01GAD}) coupled with a model of galaxy formation, where galaxy mergers are the 
primary drivers for black hole and galaxy growth and explore an additional growth channel through 
which only bulges gain mass, e.g. the disruption of stellar galactic disks in major disk mergers. 
He argues, that if the bulge growth rate from such disrupted disks is not constant with time, an 
evolution in the $M_{\bullet}$-$M_{\mathrm{bulge}}$-relation can occur.
Furthermore, \citet{Robertson06b} find from simulations of galaxy mergers, that the $M_{\bullet}$-$\sigma_*$
-relation shows a slight redshift evolution towards higher black hole masses for a given $\sigma_*$ at higher redshifts, 
but they predict no evolution for the $M_{\bullet}$-$M_{\mathrm{Bulge}}$-relation.
However, \citet{Hopkins07} using again simulations of major galaxy mergers show that high redshift black holes 
will be more massive at a fixed bulge mass than expected from the present-day relation. They find an evolution 
towards lower black hole to bulge mass ratios with cosmic time which is driven by the fact that disks (merger 
progenitors) have characteristically larger gas fractions at high redshifts. We want to point out that these studies (\citealp{Croton06, Robertson06b, Hopkins07}) are consistent with each other, they are only focusing on different aspects of the evolution of the relation.
 However, these methods do not provide a sufficient explanation for the observed under-massive black hole masses at high redshifts, since these methods find to have black holes mainly lying above the median $M_{\bullet}$-$M_{\mathrm{Bulge}}$-relation at high redshift evolving towards the relation with decreasing redshift. However, they also predict that the black holes go through a rapid growth phase before they end up above the relation; these objects might be under-massive. Furthermore, if a black hole at high z is already above the median $M_{\bullet}$-$M_{\mathrm{Bulge}}$-relation it will not end below the median relation at z=0 if 
only assuming merging and not gas accretion.
Therefore, an alternative explanation for the observed high \textit{and} low $M_{\bullet}/M_{\mathrm{bulge}}$-ratio at high redshifts 
could be the existence of a larger intrinsic scatter in black hole mass, even assuming no evolution of 
the mean relation with cosmic time, what would be in agreement with \citet{Lauer07}.

In this paper we address the question: how does the intrinsic 
scatter in black hole mass evolve and change with time assuming that black holes grow only via mergers? 
An answer to this question is also important with respect to similarities and differences between 
the observed scatter of black hole masses and predictions from theoretical models. 
\citet{Malbon}, using semi-analytic modelling, find that the present day scatter in black hole mass 
decreases significantly with increasing black hole mass. This is in contrast to observations using 
the full samples of e.g. \citet{Gebhardt00}, \citet{Tremaine02}, and \citet{Gueltekin09}. Here, the 
scatter appears to be independent of black hole mass. In particular at the high mass end the observed 
scatter is much larger than the model predictions (see Fig. \ref{Comparison}).
However, \citet{Gueltekin09} 
demonstrate that the scatter for non-elliptical galaxies (typically at lower masses) is larger than 
for elliptical galaxies. This is, at least qualitatively, in agreement with the predictions from \citet{Malbon}. 
However, in particular at the high mass end the observed scatter is significantly larger than the model 
predictions (Fig. \ref{Comparison}). 

So far the time evolution of the scatter in black hole mass has not been investigated in detail. 
\citet{Pengrandom} deals with the evolution of the scatter, but he mainly focuses on the aspect 
of how the present day $M_{\bullet}$-$M_{\mathrm{bulge}}$-relation can form in a simple model applying random 
merging of galaxies. He claims that the relation develops even if black holes and bulges are uncorrelated or incorrectly correlated in the beginning. This behaviour is supposed to 
result from an initially exponentially decreasing SMBH mass function where minor mergers drive the objects 
towards the observed correlation. Furthermore, the scatter in black hole mass decreases with increasing 
merger number, and according to his results the decrease in scatter is dominated by major mergers. 
However, a quantitative study of the scatter evolution was not presented, which is the main subject 
of this paper. A further major difference between our work and that of \citet{Pengrandom} is that we include - besides Monte-Carlo generated random merging scenarios - merging as it is found for dark matter haloes in large scale \textit{cosmological} simulations.

In the following we will investigate the evolution of the intrinsic scatter assuming:
\begin{itemize}
\item Simple random merging (section \ref{depletion})
\item Modified random merging (section \ref{replenishment}-\ref{Major_minor_Rep})
\item Merging in $\Lambda$CDM-simulations (section \ref{CDMmerging})
\end{itemize}
We point out that random merging does not describe a full physical evolution process according to 
currently favoured structure formation models. In principle the model follows dry merging of 
galaxies and therefore is limited to high mass galaxies, since merging at the high mass end 
is assumed to be almost dry, so that gas physics and star formation do not play an important role and 
can be neglected in the growth processes. However, an advantage of using a simple model such as random 
merging is that we can study separate effects on the scatter evolution, e.g. the influence of the 
initial mass distribution, of the merger mass-ratio or different refill-scenarios. Then we compare 
these results to merging according to currently favored structure formation models based on dark matter 
$\Lambda$CDM-simulations.
Since we know that a significant contribution to black hole growth is caused by accretion, we will discuss this issue in section \ref{conclusion}.

%*****************************************************************************************************
%*****************************************************************************************************
\section{Models for random merging}\label{Random} 
%*****************************************************************************************************
%*****************************************************************************************************

%*****************************************************************************************************
\subsection{Initial conditions}\label{Inidis}
%*****************************************************************************************************

In the following we describe two different initial distributions of bulges including black holes as a 
starting point for random merging: a log-normal distribution and a Schechter distribution of bulges.

\subsubsection{Initial log-normal distribution}
%*************************************************

\begin{figure}
\begin{center}
  \epsfig{file=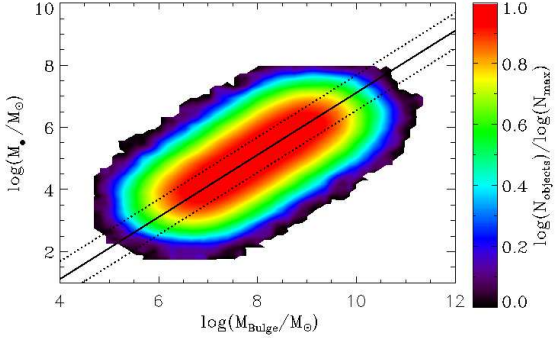, width=0.45\textwidth}
  \caption{Normalized 2D-histogram of the initial distribution with the same, log-normal distributed scatter ($\sigma_{\mathrm{ini}} = 0.6$)
in bulge and black hole mass. The observed relation is shown by the black, solid line. The black dotted lines 
indicate the  1-$\sigma$ range of the initially applied scatter in our model.}
  {\label{Inilogscatter}}
\end{center}
\end{figure}

We constructed the initial log-normal distribution by taking a uniform distribution (in the log) of bulge masses with black 
hole masses according to \citet{Haering}. Then we applied to the bulge as well as black hole masses a log-normal distributed 
scatter with a value of $\sigma = 0.5$. Having set this scatter to bulge \textit{as well as} to black hole masses results in a 
larger scatter in black hole and bulge masses with a value of $\sim 0.6$. This scatter is larger than the observed,
present-day one specified by \citet{Tremaine02} and \citet{Gueltekin09}. We have chosen this fiducial value a posteriori 
as in our most realistic $\Lambda$CDM-simulation this results in a final observed scatter value of $\sigma \sim 0.32$ 
(see section \ref{CDMmerging}). However, the scatter evolution is independent of the choice of the initial scatter value
(see section \ref{ScatterQuant}). We start with an initial distribution of $580,000$ bulges with black holes. The black 
hole masses range from $ 2.0 < \log(M_{\bullet}/M_{\odot}) <8.0$ with a mean of 
$\langle \log (M_{\bullet}/M_{\odot}) \rangle = 5.0$ and the bulges have masses of 
$ 4.7 < \log(M_{\mathrm{Bulge}}/M_{\odot}) < 11.1$ with a mean of $\langle \log (M_{\mathrm{Bulge}}/M_{\odot}) \rangle = 7.9$. 
The resulting distribution of bulges including black holes is depicted in the $2$-D histogram in Fig. \ref{Inilogscatter}. 
In this plot we have normalized the number of objects $N_{\mathrm{objects}}$ to the maximum number of objects $N_{\mathrm{max}}$ 
found in a black hole-bulge mass bin. The observed $M_{\bullet}$-$M_{\mathrm{bulge}}$-relation according to \citet{Haering} 
is plotted together with the $1$-$\sigma$-range of the applied scatter.

\subsubsection{Initial Schechter distribution}
%*******************************************************

\begin{figure}
\begin{center}
  \epsfig{file=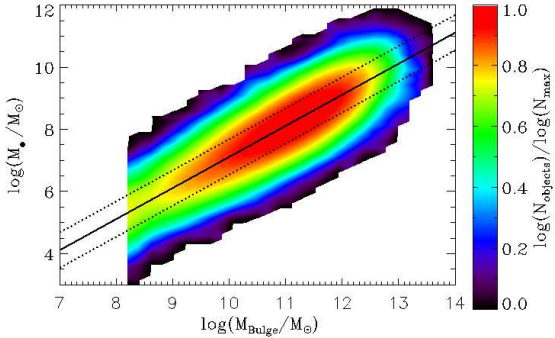, width=0.45\textwidth}
  \caption{Initial distribution based on the Schechter-fit for early-type galaxies at $z \sim 2$. An initial scatter is only applied to the black hole mass ($\sigma_{\mathrm{ini}} = 0.6$)}
  {\label{Schechter_ini_z2}}
\end{center}
\end{figure}

Observationally, we know that the mass function of bulges follows a Schechter function (e.g. \citealp{Bell03}) rather than 
a log-normal distribution. Therefore we study the scatter evolution of a Schechter-shaped initial distribution of bulge masses. 
We use a fit to the measured luminosity function (K-band magnitude) for red galaxies at redshift $z \sim 2$  according to 
\citet{Cirasuolo07} to construct our initial galaxy sample,
\begin{equation}
\Phi(M_k) = 0.4 \ln{(10)}  \cdot \Phi^* \cdot 10^{-0.4\Delta M_k(\alpha+1)} \cdot \exp{(10^{-0.4\Delta M_k})}, 
\end{equation}
with the fitting parameters
\begin{align}
\alpha & = -0.1 \nonumber\\
M_k^* & = -23.04 \nonumber\\
\Phi(10^{-3}\mbox{Mpc}^{-3}) & = 0.2 \nonumber
\end{align}
where $M_k$ are the absolute magnitudes in K-band. We convert the luminosity function into a mass function using 
the mass-to-light ratios as function of the K-band magnitude according to \citealp{Cappellari}:
\begin{align}
\frac{M}{L} & = 1.88 \cdot \left( \frac{L_k}{10^{10} \cdot L_{k,\odot}} \right) \quad \text{with}\\
M_k & = -2.5 \log{L_k} + 3.28 \nonumber
\end{align}
These mass-to-light ratios were measured for a population of ellipticals at $z = 0$ and for simplicity we assume no 
evolution with redshift. A consequence of this assumption is, that we obtain quite massive galaxies at $z=2$, although the stellar population was not evolved completely at this time. Probably, the mass-to-light ratio was smaller at higher $z$ than the present-day value leading to smaller galaxy masses. However, for our study it is sufficient to see the evolution of statistical properties, which are independent of the exact choice of the mass-to-light ratio. We scale the resulting mass distribution to a volume of $(500 \mathrm{Mpc})^3$ 
and study the evolution of $\sim 100000$ bulges with black holes, considering only bulge masses larger than 
$1.6 \times 10^{8} M_{\odot}$ ($\mathrel{\widehat{=}} \log (M_{\mathrm{Bulge}}/M_{\odot}) > 8.2 $). To keep the 
Schechter-distribution for the bulge masses we have only applied a log-normal scatter of $\sigma = 0.6$ to the 
black hole masses (see Fig. \ref{Schechter_ini_z2}). 

%*****************************************************************************************************
\subsection{Depletion scenario}\label{depletion}
%*****************************************************************************************************

For our fiducial random merging scenario (depletion case, i.e. without refilling the initial distribution with new galaxies) we use either the initial log-normal or Schechter 
distribution as described above. From the initial pool we randomly select two objects, merge them by 
adding their black hole and bulge masses and put the merged object back into the pool. In the next 
step we again merge two objects randomly, but now from the new rearranged pool. This procedure is 
repeated iteratively until, on average, every object has had one merger, i.e. only half of the initial 
objects are left over. At this point we define one merging generation to be completed.  
Then all remaining objects are considered as the initial pool for the next generation. 
Therefore, after the first generation we have $N(1) = N_{\mathrm{ini}}/2$ objects and after the $n$-th 
generation our pool is reduced to $N(n) = N_{\mathrm{ini}}/2^n$ objects. We note that in one generation some 
objects can have merged several times while others have not merged at all. Please note that if not stated otherwise, here and in the following we define the number of mergers by counting all mergers that occur for galaxies $>10^{4.7}$ M$_{\odot}$ independent of their mass ratio. 

%*****************************************************************************************************
\subsubsection{Evolution of the black hole-bulge mass relation}
%*****************************************************************************************************

\begin{figure*}
\begin{center}
  \epsfig{file=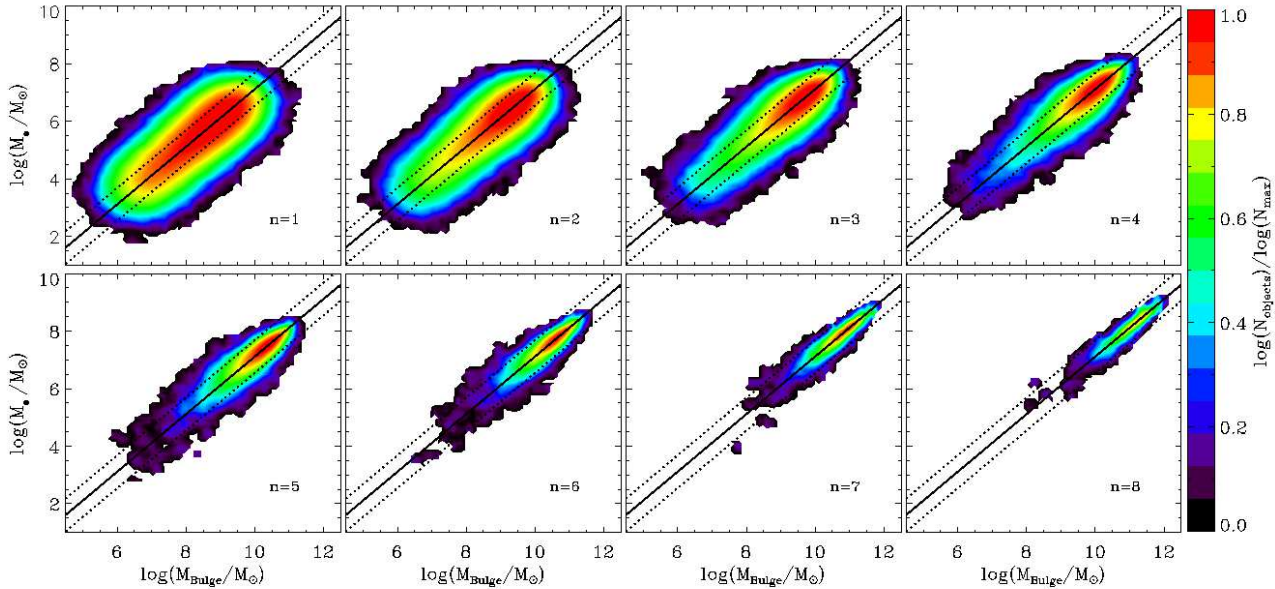, width=1.0\textwidth}
  \caption{Normalized 2D-histograms for the random merging generations $1-8$ on the basis of an initial log-normal distribution, as shown in Fig. \ref{Inilogscatter} (depletion case). The fit to the observed relation is illustrated by the black, solid line; the black dotted lines show the $1-\sigma$ range of the initially applied scatter.}
  {\label{RM_bothscatter}}
\end{center}
\end{figure*}

If we randomly merge the galaxies from the initial log-normal or Schechter distribution in Fig. \ref{Inilogscatter} 
and \ref{Schechter_ini_z2}, an important consequence of the model is, as already pointed out by \citet{Pengrandom}, 
that the sample behaves according to the \textit{central-limit-theorem} (CLT).  This theorem predicts that independent 
of the initial distribution we always converge towards a Gaussian distribution. We see this trend already after only 
one merging generation.

In Fig. \ref{RM_bothscatter} we show the evolution in the black hole-bulge mass plane 
of the log-normal distributed sample for merger generations $n=1-8$. Again, the black solid line shows 
the observed, present day, $M_{\bullet}$-$M_{\mathrm{bulge}}$-relation with the 1-$\sigma$ range of the assumed $\sigma_{\mathrm{ini}}=0.6$ initial 
scatter. The relation is conserved during all merging generations. Note, however, that here we use the \textit{same} 
initial scatter in black hole and bulge masses. In addition, the overall scatter decreases significantly with increasing merger 
generation. The low mass end of the distribution is depleted by merging whereas the high mass end is populated. We also 
see the trend that the scatter decreases more for more massive black holes and bulges than low mass systems. 

\begin{figure*}
\begin{center}
  \epsfig{file=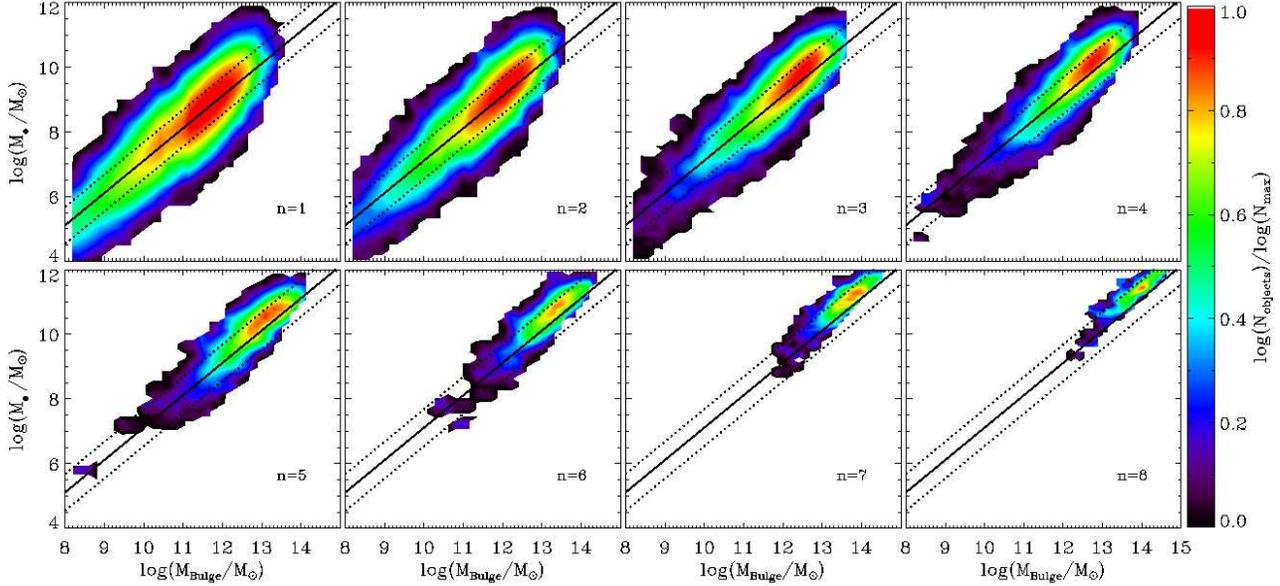, width=1.0\textwidth}
  \caption{Same as Fig. \ref{RM_bothscatter}, but for random merging with an initial Schechter-distribution resembling 
$z \sim 2$ red galaxies.}
  {\label{Schechter_merg_z2}}
\end{center}
\end{figure*}

The evolution of a randomly merging initial Schechter-distribution of bulges is shown in Fig. \ref{Schechter_merg_z2}. In this case the 
overall slope increases for small merger generations. For high merger generations, when the low mass end is depopulated, 
the slope again becomes similar to the initial slope and the relation is shifted towards larger black hole masses. The reason for the 
change in tilt as well as the shift is the different initial scatter in black hole and bulge masses. 
The shift towards larger black hole masses shows that having such initial conditions of a larger scatter in black hole 
masses than in bulge masses is quite unlikely if we want to explain the over-massive black holes at high redshift.
However, qualitatively the scatter evolution is similar to the previous case. The scatter decreases 
with increasing merging generation and a quantitive scatter estimate is presented in the next section \ref{ScatterQuant}.
We note that the initially Schechter distributed bulges evolve into a log-normal distribution for massive systems 
as a consequence of the CLT. Still, this might not be a bad approximation for massive galaxies as in principle the Schechter distribution  is a superposition of multiple Gaussians (Blanton et al. 2003). 

%*****************************************************************************************************
\subsubsection{Quantifying the scatter in the black hole mass relation}{\label{ScatterQuant}}
%*****************************************************************************************************

We characterize the scatter as the $\sigma$ in a log-normal-like distribution:

	\begin{eqnarray}\label{Gaussian}
	f(x) = \frac{1}{\sigma \sqrt{2\pi}} \cdot e^{-\frac{(x-\mu)^2}{2\sigma ^2}}\\
	\mathrm{with} \quad x = \log(M_{\bullet}) \quad \mathrm{and} \quad \mu = \langle\log(M_{\bullet})\rangle. \nonumber
 	\end{eqnarray}

Note that we use here the logarithm in base 10 '$\log$' instead of the natural logarithm '$\ln$'. This, however, 
only changes the normalization. We choose the '$\log(M)$' representation to be consistent with the observations 
(e.g. \citealp{Tremaine02}, \citealp{Gueltekin09}). To estimate the scatter $\sigma$ for the black hole mass in the 
evolution of the $M_{\bullet}$-$M_{\mathrm{bulge}}$-relation (Fig. \ref{RM_bothscatter}) we use the following method. For each merging 
generation the bulges are divided into different mass bins. Then for each bin we construct black hole mass histograms 
which resemble normal distributions which we fit with Eq. \ref{Gaussian} to derive the scatter $\sigma$. This method 
is consistent with the scatter determination in observations (\citealp{Gueltekin09}). The fit is performed with a 
Lebenberg-Marquardt-algorithm which interpolates between the Gauss-Newton algorithm and the method of gradient descent 
and searches iteratively for the best fit.

\begin{figure}
\begin{center}
  \epsfig{file=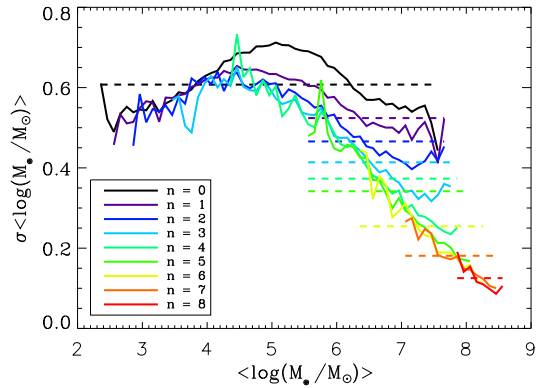, width=0.45\textwidth}
  \caption{Scatter $\sigma$ vs. mean black hole mass $\langle \log(M_{\bullet}/M_{\odot}) \rangle$ per bin for 
different merging generations $n$ in the random merging model (depletion case, initial log-normal distribution). 
$n=0$ is the initial distribution. The average values of $\sigma$ for black hole masses higher than $10^5 M_{\odot}$ within 
one merging generation are shown by dotted lines. This shows a continuous decline in scatter with merger generation.}
  {\label{Sigma_mass}}
\end{center}
\end{figure}

Fig. \ref{Sigma_mass} shows the scatter $\sigma$  as a function of mean black hole mass
$\langle \log M_{\bullet} \rangle$ per bin for the initial log-normal distribution ($n =0$) and 
the eight subsequent merging generations ($n = 1...8$). For the initial distribution the scatter on average is $\sigma =0.6$. 
However, it is not constant with mass as we have applied a log-normal distributed scatter to the bulge masses as well as 
the black hole masses. Higher merger generations show a decreasing average scatter for black hole masses larger than $10^5 M_{\odot}$(indicated by the dashed lines) as well as a decrease in scatter for larger 
black hole masses within a merging generation. This indicates a strong correlation between the scatter, 
the black hole mass, and the merging generation.

\begin{figure}
\begin{center}
  \epsfig{file=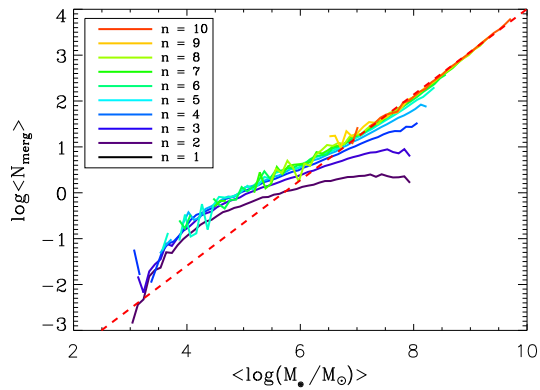, width=0.45\textwidth}
  \caption{Black hole mass $ \langle \log (M_{\bullet}/M_{\odot}) \rangle$ as a function of the mean number of merger $\langle N_{\mathrm{merg}} \rangle$ for different merging generations $n$ (depletion case, initial log-normal distribution). For higher generations $n$, $\log \langle N_{\mathrm{merg}} \rangle$ correlates with $\langle \log (M_{\bullet}/M_{\odot}) \rangle$. The linear fit is shown by the dashed, red line.}
  {\label{Linear_mass_merger}}
\end{center}
\end{figure}

\begin{figure}
\begin{center}
  \epsfig{file=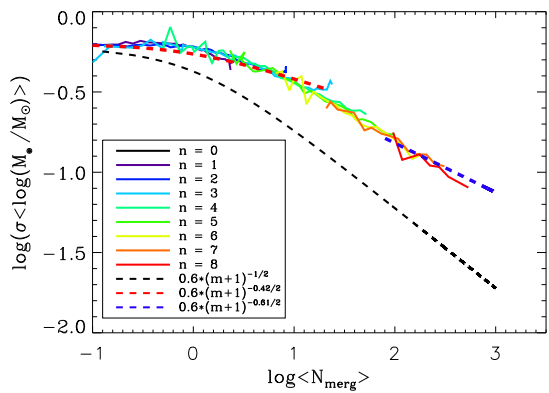, width=0.45\textwidth}
  \caption{Scatter $\log(\sigma)$ vs. mean number of mergers $\log(\langle N_{\mathrm{merg}} \rangle)$ 
for different merging generations $n$ (depletion case, initial log-normal distribution). The black 
dashed line shows the analytic solution according to the CLT for an initial normal distribution. 
The blue dashed line is a fit to the scatter for $\langle N_{\mathrm{merg}}\rangle > 100$, the blue one for $\langle N_{\mathrm{merg}}\rangle < 10$.} 
  {\label{Sigma_merger}}
\end{center}
\end{figure}

In Fig. \ref{Linear_mass_merger} we show the dependence of the mean number of mergers $\langle N_{\mathrm{merg}} \rangle$ 
on the black hole mass $\langle \log (M_{\bullet}) \rangle$ for the $8$ different merging generations indicated by the colored lines.  
At the high mass end and for merger generations $n \ge 7$ we find convergence towards a linear relation between black hole mass and mean number of mergers. 
The linear relation in Fig. \ref{Linear_mass_merger} (red, dashed line) is given by:
\begin{eqnarray}
\quad \quad \quad \quad \log \langle N_{\mathrm{merg}} \rangle =a \cdot \left\langle \log (M_{\bullet}/M_{\odot}) \right\rangle + b\\
\quad \quad \quad \quad \mathrm{with} \quad a = 0.93 \quad \mathrm{and}  \quad b = -5.32. \nonumber
\end{eqnarray}
This relation suggests, that after several merging generations $n$, we can predict how many mergers a black hole of a certain 
mass must have experienced on average. E.g. a typical supermassive black hole of $10^8 M_{\odot}$ had about 100 mergers, taking into account all mergers which an object have had during its evolution (i.e. not only mergers in the main branch, but all progenitors in the tree since the first merging generation).
Most importantly, in Fig. \ref{Sigma_merger} we plot the scatter $\sigma$ versus the mean number of mergers 
$ \langle N_{\mathrm{merg}} \rangle$ within one bulge mass bin for the different merging generations (indicated by different colors). 
The decrease of scatter with increasing merger number is an important consequence of the CLT. Hence, we can derive from the 
CLT an analytic expression for the scatter decrease for an initial normal distribution as a function of the 
merging generation $n$ (\citealp{Pengrandom}): 
	\begin{equation}\label{ana_1}
	\sigma_{\mathrm{merg}}(n) \approx \sigma_{\mathrm{ini}} \cdot 2^{-n/2},
	\end{equation}
where $n$ is the generation number and $\sigma_{\mathrm{ini}}$ is the initial scatter applied to black hole and bulge masses. 
The mean number of mergers $m$ of objects within one merging generation as a function of the generation number $n$ can be written as 
	\begin{equation}\label{connect} 
	m = 2^n - 1,
	\end{equation}
Note, that for the calculation of the mean number of mergers we consider all merger events, which galaxies had undergone until this merging generation.
With help of Eq. \ref{connect} we can rewrite Eq. \ref{ana_1} to get the scatter $\sigma$ as a function of the mean number of mergers $m$:
	\begin{equation}\label{ana_2}
	\sigma_{\mathrm{merg}}(m) \approx \sigma_{\mathrm{ini}} \cdot (m+1)^{-1/2}. 
	\end{equation}

\begin{figure}
\begin{center}
  \epsfig{file=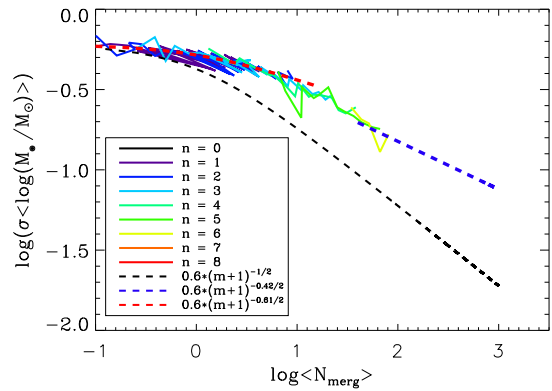, width=0.45\textwidth}
  \caption{Same as in Fig. \ref{Sigma_merger}, but based on the initial Schechter-distribution at $z \sim 2$.  }
  {\label{Sigma_merger_z2}}
\end{center}
\end{figure}
This analytic expression is depicted in Fig. \ref{Sigma_merger} by the black dashed line. Thereby we have made the assumption 
that $m$ (= mean number of mergers \textit{per generation}) $\approx \langle N_{merg} \rangle$ (= mean number of mergers \textit{per bulge mass bin} for 
one generation), which is a good approximation especially for high-mass objects. Please note, that therefore we will not distinguish between $m$ and $\langle N_{merg} \rangle$.  
In comparison to merger results, 
the analytic solution exhibits a stronger decrease in scatter. This is due to the fact that the CLT prediction assumes an initial normal 
distribution \citep{Pengrandom} whereas our sample has a log-normal distribution and therefore converges towards a different scatter. 
The red and blue dashed lines in Fig. \ref{Sigma_merger} are a fit to the random merging data assuming a fitting formula similar to Eq. \ref{ana_2}
with the exponent $a$ as a free parameter,  
	\begin{equation}\label{ana_3}
	\sigma_{\mathrm{merg}}(m) \approx \sigma_{\mathrm{ini}} \cdot (m+1)^{-a/2}.
	\end{equation}
For the mass range of convergence, $\sim m > 100$, $a = 0.61 \pm 0.02$ (blue dashed line). The exponent $a$ can therefore be used as a 
measure for the strength of the scatter decrease. For small merger numbers ($0 < m < 20$) we obtain a weaker scatter decrease with a value of $a = 0.42 \pm 0.02$ (red dashed line). This qualitatively different behaviour of the scatter decrease depending on the merger number range will be explained in section \ref{major_minor}. If we vary the 
initial scatter in black hole and bulge mass, we obtain the same strength of scatter decrease within the errors, i.e. $a$ remains unchanged. We have tested this for two different initial scatter values $\sigma_{\mathrm{ini}} = 0.40$ and $0.83$. Even if the value to which the scatter converges varies, the strength of the scatter decrease is the same ($a \sim 0.60 \pm 0.02$) in the limit of large $m$.

\begin{figure*}
\begin{center}
  \epsfig{file=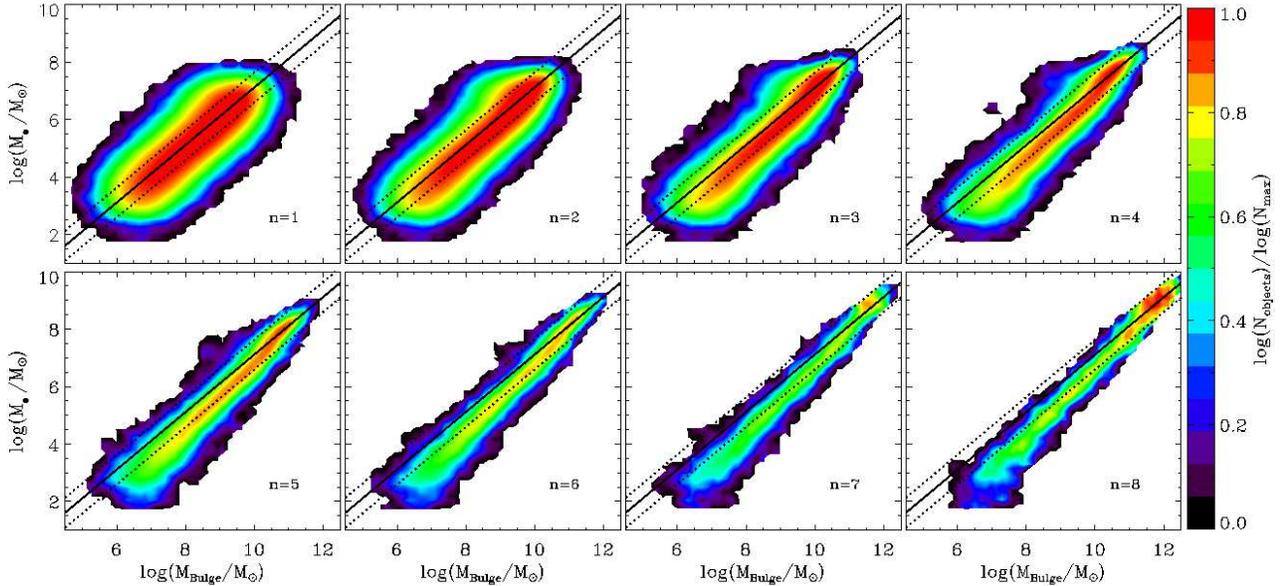, width=1.0\textwidth}
  \caption{Same as Fig. \ref{RM_bothscatter}, but galaxies had undergone only major mergers in the random merging 
model (depletion case).}
  {\label{Major}}
\end{center}
\end{figure*}

\begin{figure*}
\begin{center}
  \epsfig{file=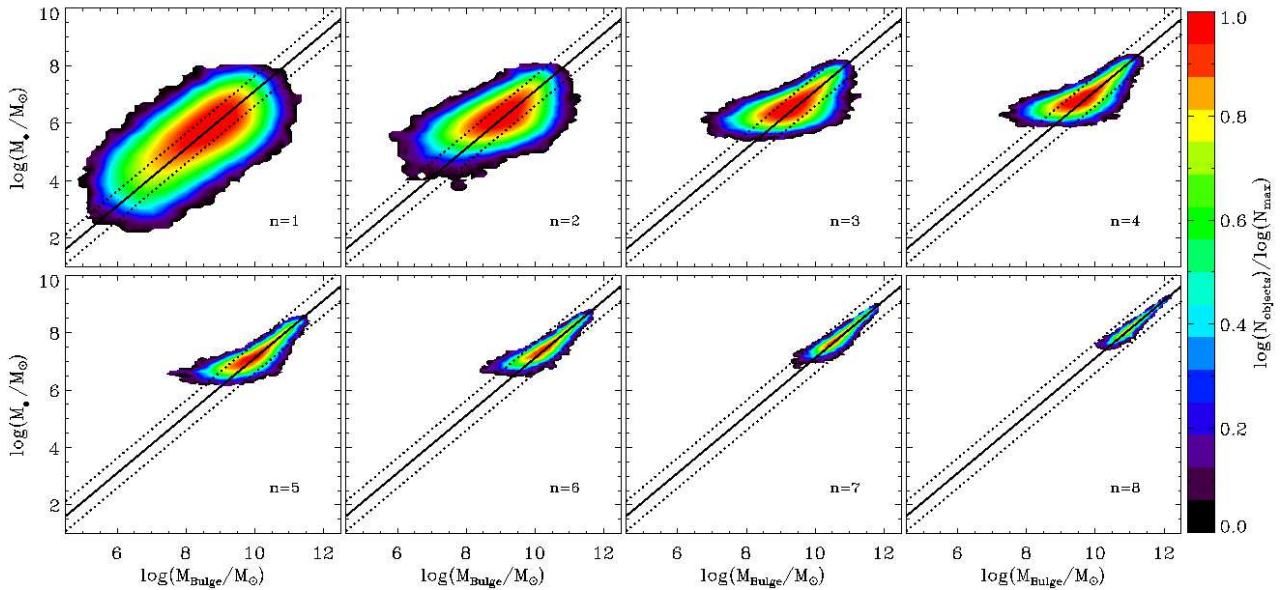, width=1.0\textwidth}
  \caption{Same as Fig. \ref{RM_bothscatter}, but galaxies had undergone only minor mergers in the random merging model 
(depletion case).}
  {\label{Minor}}
\end{center}
\end{figure*}

In Fig. \ref{Sigma_merger_z2} we present the same scatter quantification for a more realistic initial Schechter distribution. 
Assuming convergence for  $\sim m > 50$ or rather the small merger region ($0 < m < 10$) we get a similar value for $a=0.61 \pm 0.02$ or rather $a = 0.42 \pm 0.02$ (see Eq. \ref{ana_3}) 
as for an initial log-normal distribution indicating that the strength of 
the scatter decrease is \textit{weakly} dependent on the exact choice of the initial distribution. 
Again, varying the initial scatter in black hole and bulge mass does not influence the value $a$ in the exponent for large $m$.

%*****************************************************************************************************
\subsubsection{Difference between major and minor mergers}\label{major_minor}
%*****************************************************************************************************

\begin{figure}
\begin{center}
  \epsfig{file=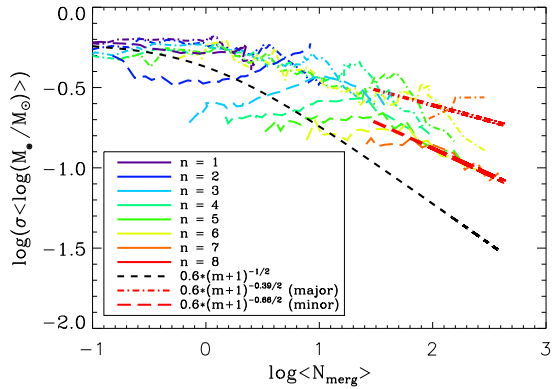, width=0.45\textwidth}
  \caption{Scatter evolution for the depletion case as a function of mean number of merger for galaxies, which had 
either only major (dotted-dashed lines) or rather only minor mergers (dashed lines).}
  {\label{ScatterMajorMinor}}
\end{center}
\end{figure}

According to \citet{Pengrandom} there is a difference for objects with only major or only minor mergers. He claims 
that major mergers exhibit a stronger central-limit tendency leading to a stronger decrease of the scatter whereas 
minor mergers are mainly responsible for evolving a linear relation between bulge and black hole masses even if 
they are uncorrelated in the beginning. Going beyond the qualitative estimates in 
\citet{Pengrandom}, we present a quantitative analysis of the scatter evolution for major and minor mergers.
We use the following definitions:
\begin{align}
 \text{Major merger:} \quad M_1/M_2 & \le 4, \quad M_1 > M_2 \label{major}\\
 \text{Minor merger:} \quad M_1/M_2 & \ge 10, \quad M_1 > M_2 \label{minor}
\end{align}
The definition for major mergers is consistent with \citet{Pengrandom}.
In Figs. \ref{Major} and \ref{Minor} we show the evolution of the black hole-bulge mass relation if we only allow major 
or minor galaxy mergers, respectively, for an initial log-normal distribution. In both cases the relation is conserved, however, by definition
for minor mergers the low and intermediate mass range is depleted. The corresponding scatter quantification is shown in Fig. 
\ref{ScatterMajorMinor}. Indeed, there is a difference between major and minor mergers; but in  
contrast to the results of \citet{Pengrandom} in the depletion case we obtain a stronger decrease of the scatter for minor mergers ($a = 0.66$) than for major mergers ($a = 0.39$).
 This result at first glance seems to contradict the expectations of the CLT. However, it is important to note, that we here consider the change of the scatter of black hole masses within individual bulge mass bins, and not over the whole population of bulge masses. During a merger generation a bulge mass bin has a constant influx and outflux of bulges due to mergers, which modifies the scatter behaviour with respect to the CLT. We have investigated the scatter behaviour for major and minor mergers in the over-all distribution of bulges. We find that when forcing galaxies to undergo only major mergers the scatter even {\textit{increases}}, whereas for minor mergers we again obtain a scatter decrease as expected from the CLT. This shows that major mergers are a very strong constraint leading to a violation of the CLT principle and causing the slower scatter decrease in smaller mass bins. 
With this behaviour we can deduce an explanation for the weak scatter decrease in the small merger number range and the stronger decrease in the limit of large merger number: the probability for having minor mergers is higher at the high mass end than for the low mass end, where major mergers dominate.

%*****************************************************************************************************
\subsection{Replenishment scenario}\label{replenishment}
%*****************************************************************************************************

\begin{figure*}
\begin{center}
  \epsfig{file=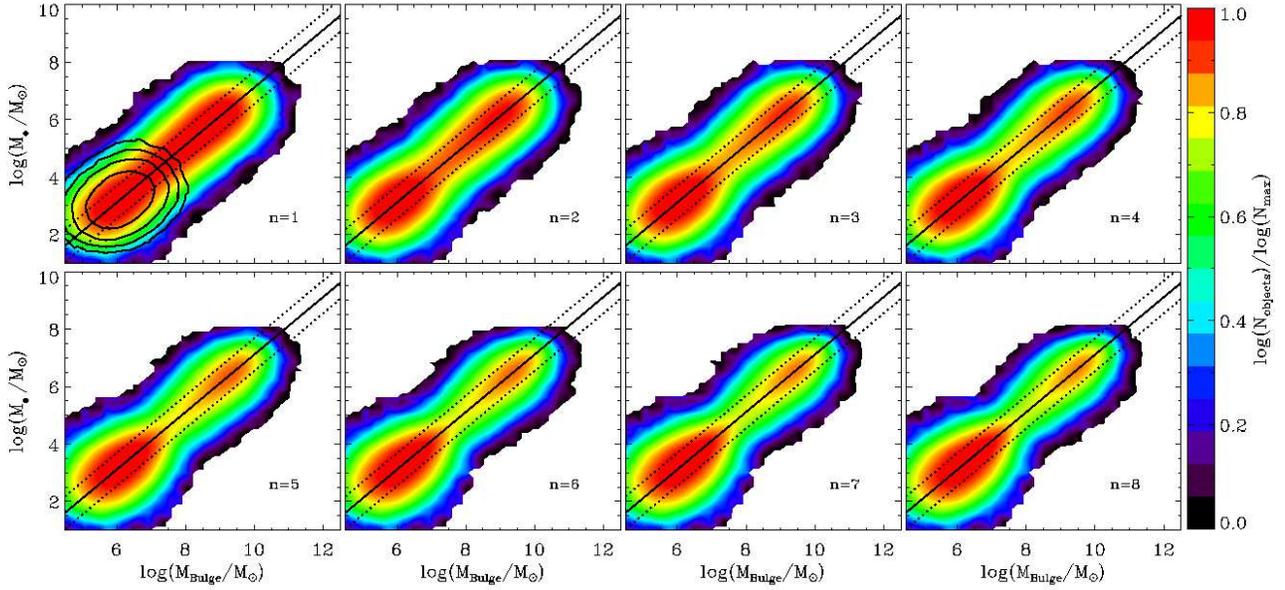, width=1.0\textwidth}
  \caption{Same as Fig. \ref{RM_bothscatter}, but for random merging (using an initial log-normal distribution) in the replenishment scenario 
with a refill-ratio of 1:1 and a low mass refill pool. The black contours in the first 
merging generation indicate the distribution of the unchanged refill-pool.}
  {\label{Repl_evol}}
\end{center}
\end{figure*}

\begin{figure*}
\begin{center}
  \epsfig{file=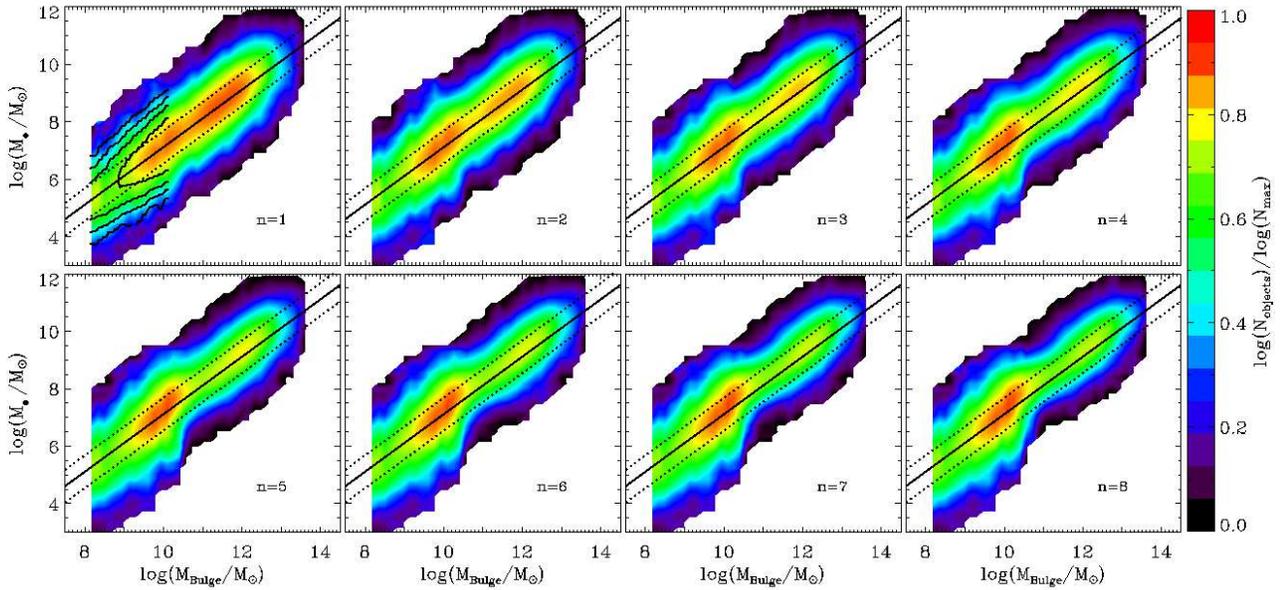, width=1.0\textwidth}
  \caption{Same as Fig. \ref{Repl_evol}, but for random merging (using an initial Schechter distribution) in the replenishment 
scenario with a refill-ratio of 1:1 and a refill pool of bulges with masses $ 10^8 M_{\odot} < M_{\mathrm{bulge}} <  10^{10} M_{\odot} $. 
The black contours in the first merging generation indicate the distribution of the refill-pool.}
  {\label{Repl_evol_Schechter}}
\end{center}
\end{figure*}
We know from observations as well as from simulations that new galaxies form during the structure formation process. To make our simple 
model more realistic we now consider different replenishment scenarios. We assume again a certain initial number of objects 
$N_{\mathrm{ini}}$ with a log-normal or a Schechter distribution and perform the same iterative random merging procedure as 
described in section \ref{depletion}. However, we now refill the pool after merger events with new objects from an external unchanged 
reservoir. We define the refill-ratio $N_{\mathrm{new}}/N_{\mathrm{event}}$ to be the number of objects added from the refill pool 
$N_{\mathrm{new}}$ per number of merger events $N_{\mathrm{event}}$ within one merging generation. The definition of one merging 
generation is the same as before but after the $n$-th merging generation the sample always contains more than $N(n) = N_{ini}/2^n$ 
objects depending on the refill-ratio $N_{\mathrm{new}}/N_{\mathrm{event}}$. At first we consider a refill-ratio of 
$N_{\mathrm{new}}/N_{\mathrm{event}} = 1$, i.e. for each merger event one new object is added randomly from the refill pool and the total number 
of objects in the sample stays constant. We also consider a refill-ratio of $N_{\mathrm{new}}/N_{\mathrm{event}} = 1/3$, i.e. one new 
object for every \textit{three} events, motivated by $\Lambda$CDM simulations (see section \ref{CDMmerging}). 

For the initial log-normal distribution, we consider a refill pool identical to the initial distribution as well as a pool 
of smaller mass galaxies with mean black hole masses $\langle \log (M_{\bullet}/M_{\odot}) \rangle \sim 3.3$ and the same initial 
scatter. For the initial Schechter distribution we use either the initial Schechter distribution itself as a refill-pool or we use the 
Schechter distribution containing only bulges at the low mass end with $m_{\mathrm{bulge}} = 1.58 \times 10^8 - 1.58 \times 10^{10} M_{\odot}$. The cases with 
smaller refill pools (lower galaxy masses) allow a more realistic comparison to the $\Lambda$CDM-simulations presented 
in section \ref{CDMmerging}. In total we have four different refill-scenarios which will be investigated in the following:
\begin{enumerate}\renewcommand{\labelenumi}{\arabic{enumi}.}
\item Refill-ratio 1:3 \& initial mean (\textit{Ini 1:3})
\item Refill-ratio 1:3 \& small mean (\textit{Small 1:3})
\item Refill-ratio 1:1 \& initial mean (\textit{Ini 1:1})
\item Refill-ratio 1:1 \& small mean (\textit{Small 1:1})
\end{enumerate}

%*****************************************************************************************************
\subsubsection{Evolution of the black hole-bulge mass relation}
%*****************************************************************************************************

A general feature of all replenishment models is that the scatter in the $M_{\bullet}$-$M_{\mathrm{Bulge}}$-relation 
is again reduced with increasing merger number. However, compared to the depletion scenario, more merger generations 
are needed to reduce the scatter by the 
same amount. In other words, for the same merger generation the scatter decrease is weaker as new objects with 
a larger initial scatter are added. For small refill pools and large refill-ratios we find an interesting feature. Fig. \ref{Repl_evol} 
and Fig. \ref{Repl_evol_Schechter} show the evolution of the relation for a refill-ratio of one and the small (low mass) refill pools 
(\textit{Small 1:1}) for an initial log-normal and a Schechter distribution. The contours in the plot of the first random merging 
generation illustrate the distribution of the unchanged refill-pool. In both cases a double peak structure emerges. 
This is a consequence of using a low mass refill pool and a high refill-ratio. The low mass peak reflects the appearance 
of new objects chosen from refill-distribution whereas the high mass peak evolves through merging from the initial 
distribution, similar to the simple case (section \ref{depletion}). This behaviour will be discussed in section \ref{CDMmerging} in more detail.

%*****************************************************************************************************
\subsubsection{Quantifying the scatter in black hole relations}\label{scatter_refill}
%*****************************************************************************************************

\begin{figure}
\begin{center}
  \epsfig{file=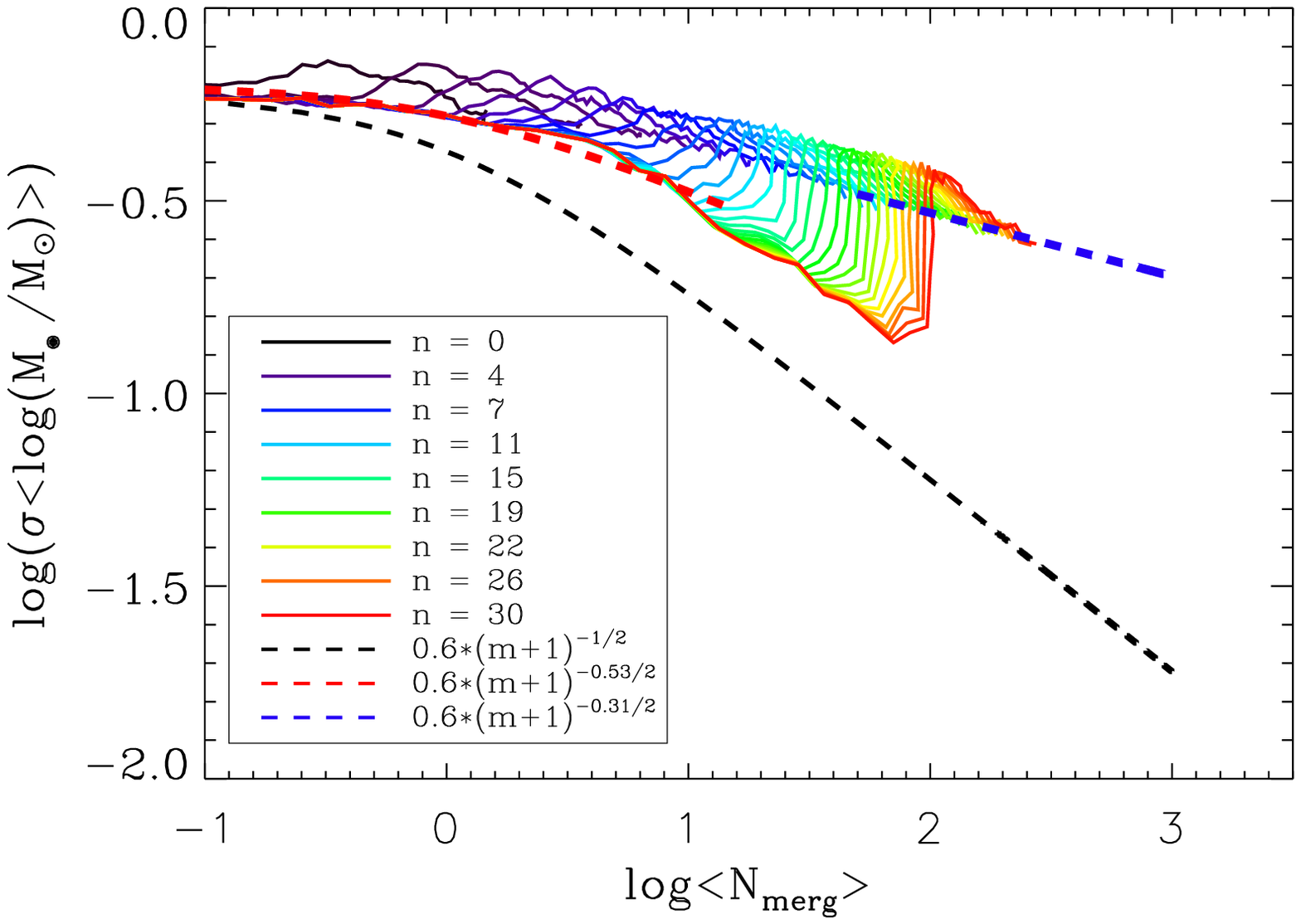, width=0.45\textwidth}
  \caption{Same as in Fig. \ref{Sigma_merger}, but for a replenishment scenario with a refill-ratio of $1$, a refill-pool with a small mean 
and an initial \textit{log-normal} distribution.The 'spike' near $\log\langle N_{\mathrm{merg}} \rangle = 2$ is due to the bimodality as discussed in the text.} 
  {\label{Repl_sigma_onecase}}
\end{center}
\end{figure}

\begin{figure}
\begin{center}
  \epsfig{file=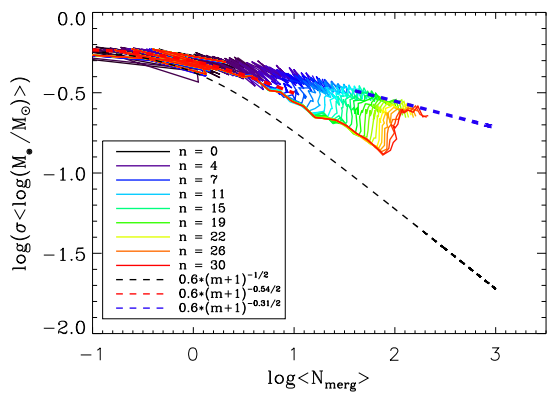, width=0.45\textwidth}
  \caption{Same as in Fig. \ref{Repl_sigma_onecase}, but for a replenishment scenario with a refill-ratio of $1$, 
a refill-pool with a small mean and an initial \textit{Schechter} distribution.}
  {\label{Repl_sigma_onecase_Schechter}}
\end{center}
\end{figure}

\begin{table}
\centering
\caption{Values of the fit parameter $a$ for scatter decrease in different random merging models for a Schechter and a log-normal initial distribution in the limit of small ($0 < m < 20$) and large merger number ($100 < m$).}
\begin{tabular}{c | c c c c c}
Ini. distr. & Depl. & Ini 1:3 & Small 1:3 & Ini 1:1 & Small 1:1\\
\hline \hline
\bf{Schechter} \\
high m & 0.61  & 0.56 & 0.50 & 0.56 & 0.32\\
small m & 0.42  & 0.38 & 0.36 & 0.34 & 0.54\\
\bf{Log-norm} \\
high m & 0.61  & 0.60 & 0.51 & 0.55 & 0.31\\
small m & 0.42  & 0.38 & 0.34 & 0.33 & 0.53\\
\hline\\
\end{tabular}
\label{Schechter-table}
\end{table}
In contrast to the depletion scenario, the replenishment models lead to a slower decrease of the scatter in black hole mass. 
The scatter quantification for the evolution of the $M_{\bullet}$-$M_{\mathrm{bulge}}$-relation in Figs. \ref{Repl_evol}  and \ref{Repl_evol_Schechter}
(refill-ratio 1, small mean and initial log-normal distribution or Schechter distribution) is shown in Figs. \ref{Repl_sigma_onecase} and \ref{Repl_sigma_onecase_Schechter}. 
The decrease of the scatter at low merger numbers originates from merging of new objects from the refill-pool mainly dominated by minor mergers 
whereas the decrease of the scatter at high merger numbers refelects merging at the high mass end (see section \ref{depletion}), mainly dominated by major mergers. We obtain a stronger scatter decrease for small merger numbers ($a = 0.53$, red dashed line) than for the large merger numbers ($a = 0.31$, blue dashed line), since the probability for having minor mergers is higher at the small merger number end.

We analyse the influence of refill-ratio and refill-pool and summarize the scatter evolution resulting from the 
four different replenishment models in table \ref{Schechter-table} for an initial log-normal and an initial Schechter distribtution. The typical errors 
are $\pm 0.02$. The larger the refill-ratio, the more slowly the scatter decreases. Keeping in 
mind that a larger refill-ratio corresponds to a larger number of new objects added per merger generation, we can explain 
this behaviour as follows. New objects from the refill-pool have a large \textit{initial} scatter in black hole mass, in contrast to 
objects after several merging generations, whose scatter has already decreased. Therefore, the more objects are added to 
the sample, the less the scatter decreases with merging generation. In addition, the lower the typical mass of objects in 
the refill pool, the more slowly the scatter decreases in the limit of large merger numbers. That means that for large merger numbers, the probability for having major mergers gets higher when using a low mass refill-pool. However, for low merger numbers the scatter decreases more rapidly with a low mass refill-pool.
 In this range, minor mergers become more likely because of the low-mass refill-pool, leading to a stronger scatter decrease than in the major merger dominated region (i.e. convergence region of large merger numbers). Altogether, a small-mass refill-pool leads to higher probability of having minor mergers in the small merger number region than at the limit of large merger numbers, where major mergers dominate.
Furthermore, as shown in table \ref{Schechter-table}, the choice of the initial distribution does not change the scatter decrease $a$ (for small as well as large $m$). We expect this from the CLT as 
the distribution quickly evolves into a normal distribution no matter which initial distribution is used.

%*****************************************************************************************************
\subsubsection{Difference between major and minor mergers}\label{Major_minor_Rep}
%*****************************************************************************************************

\begin{table}
\centering \label{valuesmajmin_repl}
\caption{Values of the fit parameter $a$ for scatter decrease in different random merging models for either only major or only minor mergers based on an initially log-normal distribution in the limit of large merger numbers ($m > 100$).}
\begin{tabular}{c c c c c c}
  & Depl. & Ini 1:3 & Small 1:3 & Ini 1:1 & Small 1:1 \\
\hline \hline
Major & 0.39 & 0.50 & 0.43 & 0.54 & 0.15\\
Minor &0.66 & 0.65 & 0.59 & 0.59 & 0.30\\
All & 0.61 & 0.60 & 0.50 & 0.55 & 0.30\\
\hline\\
\end{tabular}
\end{table}

Since \citet{Pengrandom} considered in his study a replenishment scenario with a refill ratio 
of $N_{\mathrm{new}}/N_{\mathrm{event}} = 1$ using the initial distribution as the refill pool, we will also investigate the difference between major and minor mergers for different replenishment models.
The fitted slopes for the scatter evolution for major and minor mergers are summarized in table \ref{valuesmajmin_repl} with an error of $\sim \pm 0.02$. Note that for the case refill-ratio $1:1$ with a low-mass refill-pool, we obtain a larger error of $\sim \pm 0.05$, since here the black hole mass histograms are not fitted well by a Gaussian function anymore.
As in the depletion case (see section \ref{major_minor}) we see a stronger scatter decrease for galaxies undergoing only minor mergers than for major mergers. We argue as above that this deviation from the CLT is most likely due to dividing the bulge masses in different mass bins which suffer a constant influx and outflux from bulges during each merger generation.  And furthermore, forcing galaxies to undergo only major mergers is a very strong constraint leading to a violation of the CLT.

%*****************************************************************************************************
%******************************************************************************************************8
\section{Comparison to merging in $\Lambda$CDM-Simulations}\label{CDMmerging}
%*****************************************************************************************************
%*****************************************************************************************************

So far, the influence of different idealised random merging models on the evolution of the $M_{\bullet}$-$M_{\mathrm{bulge}}$-relation 
and the corresponding scatter in black hole mass has been discussed. In this section we investigate a more complex and 
astrophysically motivated model based on merger trees from dark matter simulations following structure formation
in a $\Lambda$CDM universe.  

%*****************************************************************************************************
\subsection{Simulation setup and merger tree construction}
%*****************************************************************************************************

We have simulated a comoving periodic box with $L=100\ \mathrm{Mpc}$ box length and $512^3$ particles using 
GADGET2 code \citep{Springel01GAD}. For this simulation we assume a $\Lambda$CDM cosmology based in the 
WMAP3 (see e.g. \citet{Spergel03}) measurements with $\sigma_8 = 0.77$, $\Omega_{m}=0.26$ , 
$\Omega_{\Lambda}=0.74$, and $h=H_0/(100\ \mathrm{kms}^{-1})=0.72$ (see also \citet{Moster09} for a 
first analysis of this simulation). The simulation was started at $z=43$ and run until $z=0$ with 
a fixed comoving softening length of $2.52\ h^{-1} \mathrm{kpc}$. Starting at an expansion factor of 
$a=0.06$ we have halo catalogues for 94 snapshots until $z=0$ separated by $\Delta a =0.01$ in time. 
The mass of one dark matter particle is  $2 \times 10^8 M_{\odot}/h$. To identify dark matter haloes at 
every snapshot we use a FOF algorithm with a linking length of $b=0.2$. In a second step we extract 
subhaloes of every FOF halo using the SUBFIND algorithm \citep{Springel01GAD}. This halofinder 
identifies overdense regions and removes gravitationally unbound particles. This way we split the FOF 
group into a main or host halo and its satellite halos. The sizes and virial masses of the main halos 
(i.e. the most massive SUBFIND halos) are determined with a spherical overdensity criterion. 
The miminum halo mass is set to 20 particles ($4 \times 10^9 M_{\odot}/h$). To create the merger tree 
for the main halos (satellite halos are not considered) we connect halos between the 
snapshots as described in detail in \citet{Maulbetsch} with a few modifications which we will explain in the following. 
The branches of the trees for $z=0$ halos are constructed by connecting the halos to their most 
massive progenitors (MMP) at previous snapshots. Thereby halo $j$ with $n_j$ particles at redshift $z_j$ with the 
maximum probability $p(i,j)$ is chosen to be a MMP of halo $i$ containing $n_i$ particles at redshift $z_i$ (where $j < i$). 
The probablity $p(i,j)$ is defined as 
\begin{align}
p(i,j) & = \frac{n_{ov}(i,j)}{n_{max}(i,j)} \quad \text{with} \\
n_{ov} & = n_i(z_i) \cap n_j(z_j) \quad \text{and} \nonumber\\
n_{max}(i,j) & = \max(n_i(z_i),n_j(z_j)) \nonumber
\end{align}
Here, $n_{ov}$ is the number of particles found in both halos and $n_{max}$ is the particle number 
of the larger halo. We remove 'fake' haloes which exist only within one timestep and have no connection to any branch.
The low redshift ends of the branches are then checked for mergers. A halo $j$ is assumed to merge with 
halo $i$, if at least $50 \% $ of the particles of halo $j$ are found in halo $i$. In case of a merger 
the branches are connected. For a proper connection we apply the 'split'-algorithm \citep{Genel08} to 
prevent double or multiple counting of merger events within a tree. The 'split' algorithm was shown 
to produce more reliable merger rates which is also important for our study.

%*****************************************************************************************************
\subsection{Evolution of the relation between black hole mass and galaxy properties}
%*****************************************************************************************************

We consider two different possibilities to populate dark matter haloes with black holes and bulges. We either apply black hole masses or bulge masses directly to dark matter halos and calculate the corresponding missing quantity using the the black hole-bulge mass relation  in \citet{Haering}.
\begin{figure*}
\begin{center}
  \epsfig{file=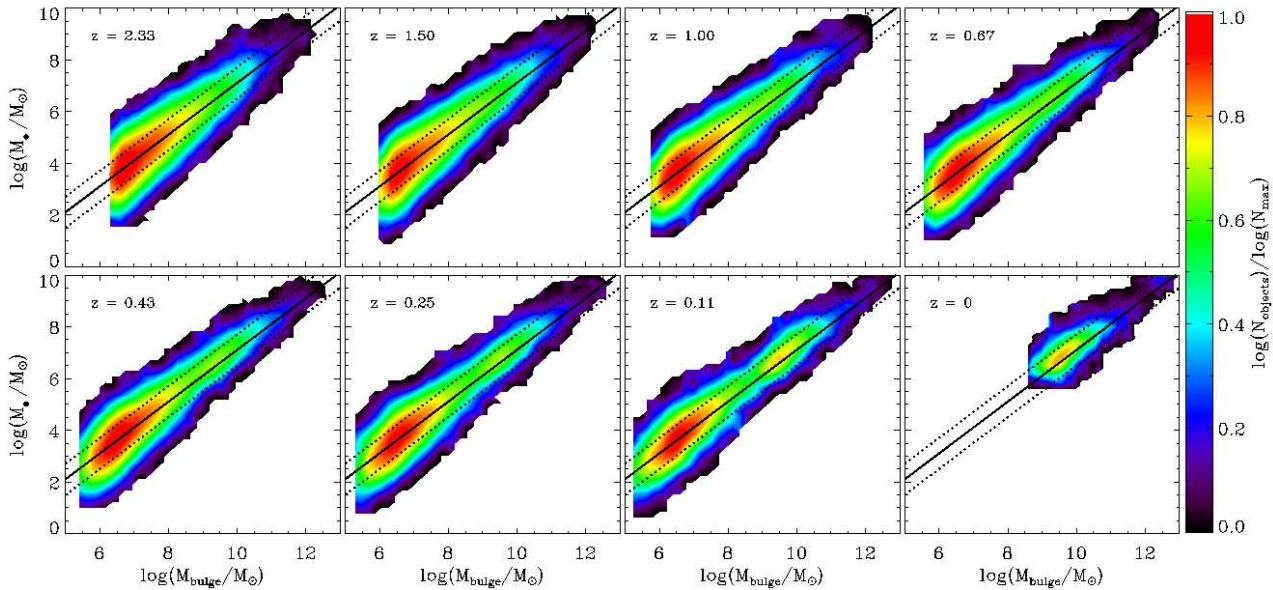, width=1.0\textwidth}
  \caption{Evolution of the $M_{\bullet}$-$M_{\mathrm{bulge}}$-relation through merging only in $\Lambda$CDM-simulations using $M_{\bullet}$-$M_{DM}$-relation.}
 {\label{Cosmo_merging_bulge}}
\end{center}
\end{figure*}

\subsubsection{Using a $M_{\bullet}$-$M_{DM}$-relation}\label{direct}

\citet{Ferr02} proposed a relation between the mass of the central black hole $M_{\bullet}$ and the mass of the dark matter halo 
$M_{\mathrm{DM}}$ of the form 
\begin{equation}\label{BH_halo_z0}
  \frac{M_{\bullet}}{10^8 M_{\odot}} \sim 0.1 \cdot \left( \frac{M_{\mathrm{DM}}}{10^{12} M_{\odot}} \right)^{1.65}.
\end{equation}
We want to point out that only the $M_{\bullet}$-$M_{\mathrm{bulge}}$-relation is directly observable and therefore assumed to be more fundamental than the $M_{\bullet}$-$M_{\mathrm{DM}}$-relation.
However, the $M_{\bullet}$-$M_{DM}$-relation has been supported by e.g. recent results of  \citet{Booth09}, where the authors obtain a similar relation between black hole and dark matter halo mass using numerical simulations (GADGET III) with self-consistent black hole growth, which are tuned to match the relations between black hole mass and galaxy stellar properties. Furthermore there are observations from \citet{Croom07, Qingjuan08, White08} which confirm the existence of a relation between black hole and dark matter halo mass using observations.
 However, this relation is only valid for $z=0$. Since we seed the black holes at higher redshift, we modify expression Eq. \ref{BH_halo_z0} in order to maintain the relation at $z=0$. We assume the same derivation as it is described in \citet{Ferr02}. Therefore we first use the virial velocities $v_{vir}$ from our simulations calculated by SUBFIND. This way we obtain $v_{vir}$ as a function of $M_{DM}$ \textit{and} redshift $z$. Assuming that $v_{vir} \approx v_c$ we use 
%But we first calculate $v_c$ as function of $M_{DM}$ \textit{and} redshift $z$ as it is presented in \citet{Somerville99}. 
then the relation between circular velocity $v_c$ and velocity dispersion $\sigma_c$ and the one between velocity dispersion $\sigma_c$ and black hole mass $M_{\bullet}$. So we can rewrite equation \ref{BH_halo_z0}:
\begin{equation}\label{BH_halo}
  M_{\bullet} = 3.12 \times \left( \frac{v_c(M_{DM},z)^{1.19}}{200\ \mathrm{km/s}} \right) ^{4.58} 10^5 M_{\odot}
\end{equation}
Note, that the relation between circular velocity $v_c$ and velocity dispersion $\sigma_c$ is derived from observations
of spiral galaxies at $z=0$. Presumably at higher redshifts dissipation of baryons has a higher influence than for $z=0$. This could lead to larger velocity dispersions and therefore also to larger black hole masses. 
According to relation \ref{BH_halo} we can populate halos with masses extracted from the dark matter simulations with 
central supermassive black holes. As we are mainly interested in massive galaxies where gas physics is assumed to 
be less important (at least at low redshifts e.g. \citealp{Dekel08,Naab06,Khochfar09}) we only investigate merger trees starting at $z=3$ 
for halos more massive than  $10^{12} M_{\odot}$ at $z=0$. 
At the high redshift end of every branch starting at $z=3$ we populate the most massive progenitors of 
the selected $z=0$ halos with black holes according to Eq. \ref{BH_halo}. 
Additionally we add a log-normal distributed scatter to the black hole masses with the $\sigma_{\mathrm{ini}} = 0.6$ 
(see section \ref{Random}). For the subsequent growth of the black holes we only take dark matter halo mergers and the corresponding 
merger of their black holes into account. 

To allow a comparison with the random merging cases shown previously, the evolution of the $M_{\bullet}$-$M_{\mathrm{bulge}}$-relation is shown in Fig. \ref{Cosmo_merging_bulge}. 
We have chosen black holes in dark matter haloes according to Eq. \ref{BH_halo} and then calculated
corresponding bulge masses by taking the median relation of \citet{Haering}. Once the dark halos have
been populated with these bulge masses we apply a scatter of sigma=0.6 to the black hole masses for
each bulge mass. We again follow the growth process only via merger events. 
The high mass end of the relation is shifted towards larger black hole masses as 
we have only applied an initial scatter to black hole masses but not to the bulge masses. Similar to the simple models investigated before again the scatter decreases with time. We also see a double peak structure 
at low redshifts $z < 0.4$ similar to the replenishment random merging model with a low mass refill pool 
(section \ref{replenishment}). This is a consequence of the conditional mass function. We only investigate 
halos under the condition that they merged into $z=0$ halos with masses larger than $10^{12} M_{\odot}$ 
(see e.g. \citealp{Somerville00}). 

\subsubsection{Using a galaxy population model}\label{Galpop}

Alternatively, to populate dark matter halos with black holes, we can use a fitting function that relates host dark halo masses to stellar masses of galaxies to assign to every 
	dark matter halo mass a galaxy stellar mass. There exist many studies which link the distribution of galaxies to that of dark matter halos (\citealp{Bosch03, Bosch07, Mandelbaum06, Moster09a, Guo09}). Here we take the fitting formula from \citet{Moster09a}. They assume that every host halo contains exactly one central galaxy and - as a constraint from the observed galaxy mass function - that the stellar mass to dark matter halo mass ratio $M_*/M_{DM}$  first increases with increasing mass, reaches a maximum and then decreases again. Hence \citet{Moster09a} adopt the following parametrization, similar to the one used in \citet{Yang03}:
\begin{equation}\label{Moster1}
\frac{M_{*}(M_{DM})}{M_{DM}} = 2 \left( \frac{M_*}{M_{DM}} \right) \left( \left( \frac{M_{DM}}{M_1} \right)^{-\beta} + \left( \frac{M_{DM}}{M_1} \right)^{\gamma} \right)^{-1}
\end{equation}
Basically, this parametrization is set to reproduce many observations, as the galaxy mass function or clustering.
Choosing a redshift parametrization for each of the parameters in Eq. \ref{Moster1}, they can predict the galaxy to dark matter mass ratio 
at any redshift:
\begin{eqnarray}
\log{M_1(z)} = \log{M_0} \cdot (z+1)^{\mu}\\
\left( \frac{M_*}{M_{DM}} \right)_0(z) = \left( \frac{M_*}{M_{DM}} \right)_{z=0} \cdot (z+1)^{\nu}\\
\gamma(z) = \gamma_0 \cdot (z+1)^{\gamma_1}\\
\beta(z) = \beta_1 \cdot z + \beta_0
\end{eqnarray} 
with $\log M_0 = 11.88$, $\mu = 0.019$, $(M_*/M_{DM})_{z=0} = 0.0282$, $\nu = -0.72$, $\gamma_0 = 0.556$, $\gamma_1 = -0.26$, $\beta_0= 1.06$ and $\beta_1 = 0.17$.

\begin{figure*}
\begin{center}
  \epsfig{file=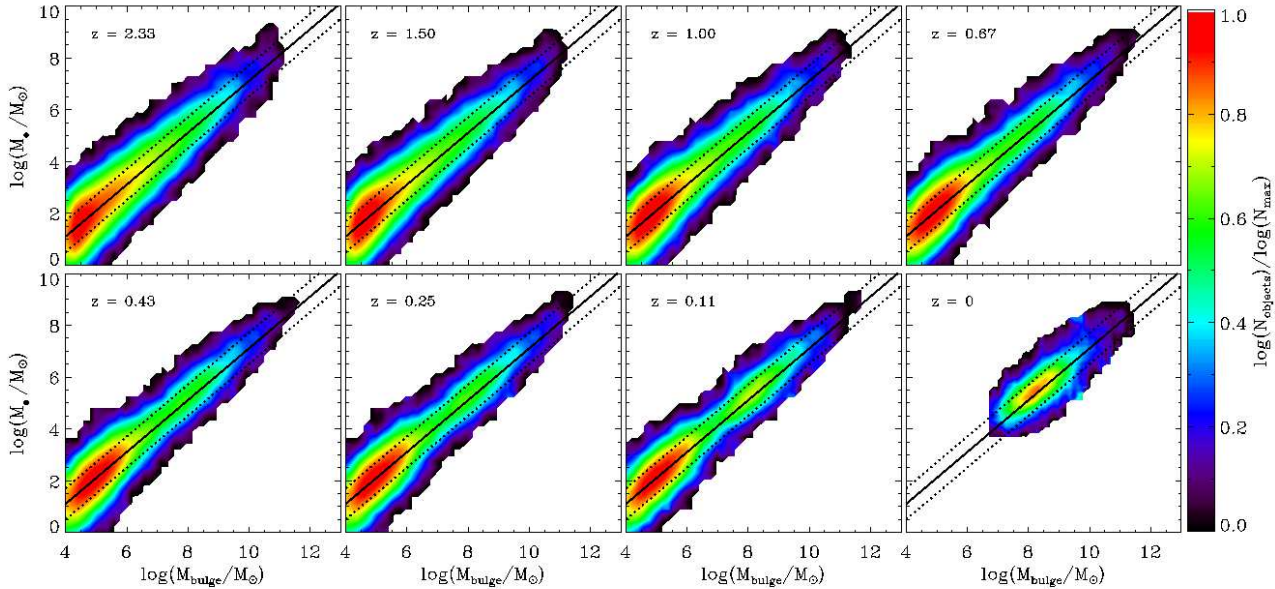, width=1.0\textwidth}
  \caption{Evolution of the $M_{\bullet}$-$M_{\mathrm{bulge}}$-relation through merging only in $\Lambda$CDM-simulations using a galaxy population model (\citealp{Moster09a}).}
 {\label{Cosmo_merging_bulge_Ra}}
\end{center}
\end{figure*}
To populate the galaxies with black holes, we assume for simplicity that all stars are in the spheroidal component of the galaxy ($M_* \approx M_{bulge}$). Using the $M_{\bullet}$-$M_{bulge}$-relation (\citealp{Haering}) we apply to each galaxy mass a black hole mass with the same initial scatter as already used before ($\sigma = 0.6$). If we then take into account the growth of black holes and galaxies through merging according to our $\Lambda$CDM-simulations, we obtain an evolution of the $M_{\bullet}$-$M_{\mathrm{bulge}}$-relation as shown in Fig. \ref{Cosmo_merging_bulge_Ra}. Again, we see the same effect as already described in subsubsection \ref{direct}: a decreasing scatter with time together with an evolving double peak structure.

%*****************************************************************************************************
\subsection{Quantifying the scatter in the black hole mass relation}
%*****************************************************************************************************

\begin{figure}
\begin{center}
  \epsfig{file=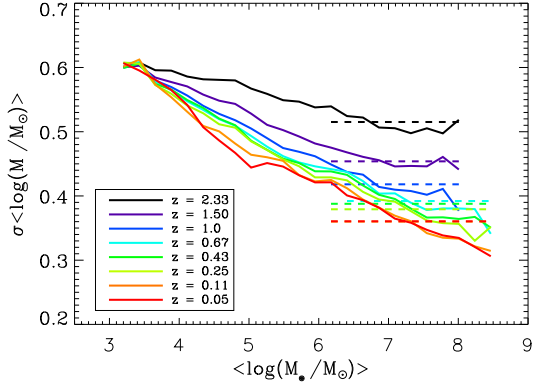, width=0.45\textwidth}
  \caption{Scatter in black hole mass versus mean black hole mass $\langle \log (M_{\bullet}/M_{\odot}) \rangle$ 
at different redshifts (different colors) in the $\Lambda$CDM-simulation assuming black hole seeding according to the $M_{\bullet}$-$M_{DM}$-relation. The horizontal lines indicate the average scatter for black holes more massive than $\approx 10^6M_{\odot}$. 
The scatter for massive black holes continously decreases towards lower redshifts.}
 {\label{Cosmo_scatter_Mass}}
\end{center}
\end{figure}

\begin{figure}
\begin{center}
  \epsfig{file=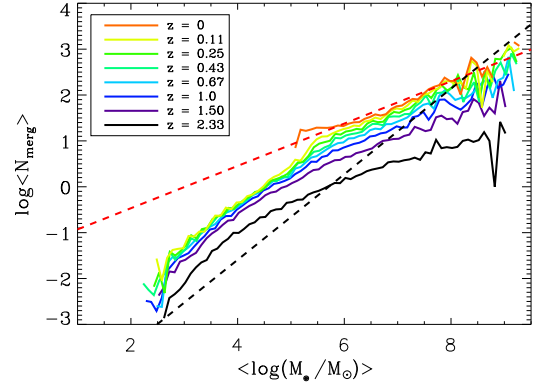, width=0.45\textwidth}
  \caption{Relation between mean black hole mass $\langle \log (M_{\bullet}/M_{\odot}) \rangle$ and mean number of 
mergers $\log \langle N_{\mathrm{merg}}  \rangle$ at different redshifts (different colors) in the $\Lambda$CDM-simulation assuming black hole seeding according to the $M_{\bullet}$-$M_{DM}$-relation. The best fit at z=0.05 
is indicated by the red dashed line. For comparison the dashed black line indicates the best fit for the random merging depletion model 
(Fig. \ref{Linear_mass_merger}).}
 {\label{Cosmo_Linear}}
\end{center}
\end{figure}

For the evolution of the $M_{\bullet}$-$M_{\mathrm{bulge}}$-relation shown in Fig. \ref{Cosmo_merging_bulge} (the model based on the relation between dark halo mass and and black hole mass) we 
performed the same scatter quantification as before for the different random merging models. 
Fig. \ref{Cosmo_scatter_Mass} shows the evolution of the scatter in black hole mass. At the high mass end, which we are focusing on here, 
the scatter decreases towards lower redshift. Furthermore - similarly to random merging - in Fig. \ref{Cosmo_Linear} 
we find convergence towards a linear relation between the logarithm of the mean number of mergers and the logarithm of black hole mass 
at low redshift. We fit this relation according to  
\begin{eqnarray}
\quad \quad \quad \quad \log \langle N_{\mathrm{merg}} \rangle = a \cdot \langle \log (M_{\bullet}/M_{\odot}) \rangle + b\\
\quad \quad \quad \quad \mathrm{with} \quad a=0.46 \quad \mathrm{and} \quad b = -1.39. \nonumber
\end{eqnarray}
shown by the red dashed line in Fig. \ref{Cosmo_Linear}. For comparison the dashed black line illustrates the random merging case (depletion).
As expected lower mass black holes ($M_{\bullet} < 10^8 M_{\odot}$) experience more mergers than predicted for the simple depletion case, whereas at the high mass end galaxies undergo less mergers than in the depletion model. 
\begin{figure}
\begin{center}
  \epsfig{file=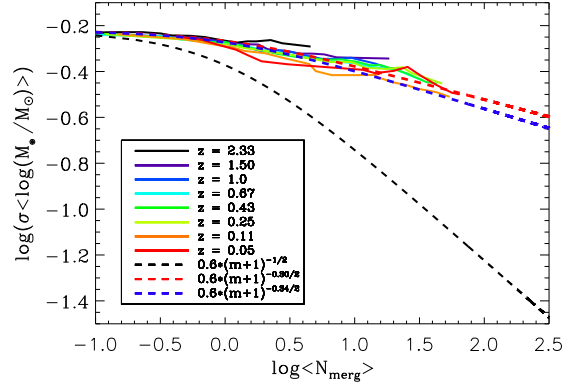, width=0.45\textwidth}
  \caption{Scatter $\log(\sigma)$ as a function of mean number of mergers $\log\langle N_{\mathrm{merg}} \rangle$ for different time steps in the $\Lambda$CDM-simulation assuming black hole seeding according to the $M_{\bullet}$-$M_{DM}$-relation. The blue dashed line shows the fit to the best-matching random merging model (refill-ratio $1/3$ and low-mass refill-pool), the black dashed line corresponds to the analytic case.}
 {\label{Cosmo_scatter_Merger}}
\end{center}
\end{figure}

Finally the scatter $\sigma$ is plotted as a function of the mean number of mergers $\langle N_{\mathrm{merg}} \rangle$ in Fig. \ref{Cosmo_scatter_Merger}. 
We find the same qualitative  behaviour as for all random merging models. The scatter decreases with increasing number of merger events. 
However, the scatter decrease in $\Lambda$CDM-merging is weaker. The fit parameter $a = 0.30 \pm 0.03$ is smaller than for most of the random merging models at the limit of large merger numbers. Note that for the case where we assign black holes to dark matter halos using the galaxy population model, as described in subsection \ref{Galpop}, we get the same qualitative scatter behaviour. The only difference is that because there are fewer objects at the high mass end, the resulting scatter value is connected with a larger error than for the black hole population described in subsection \ref{direct}.
In comparison to the best-matching 
random merging scenario - a refill-ratio of $1/3$ and a low-mass 
refill-pool - $\Lambda$CDM-merging leads to a clear difference in the scatter decrease ($a = 0.51 \pm 0.02$). However, in the case of $\Lambda$CDM-merging we have only a maximum number of mergers of $\sim 60$. Therefore, the scatter decrease in CDM-merging is consistent with the best-matching random merging model only in the low merger range ($a = 0.34 \pm 0.02$), indicated by the blue, dashed line. 
This shows that the scatter evolution in  $\Lambda$CDM-merging can be well approximated - quantitatively and qualitatively - by the simple model of random merging without any structure formation model.

%*****************************************************************************************************
\subsection{Evolution of the black hole mass function}
%*****************************************************************************************************

\begin{figure*}
\begin{center}
  \epsfig{file=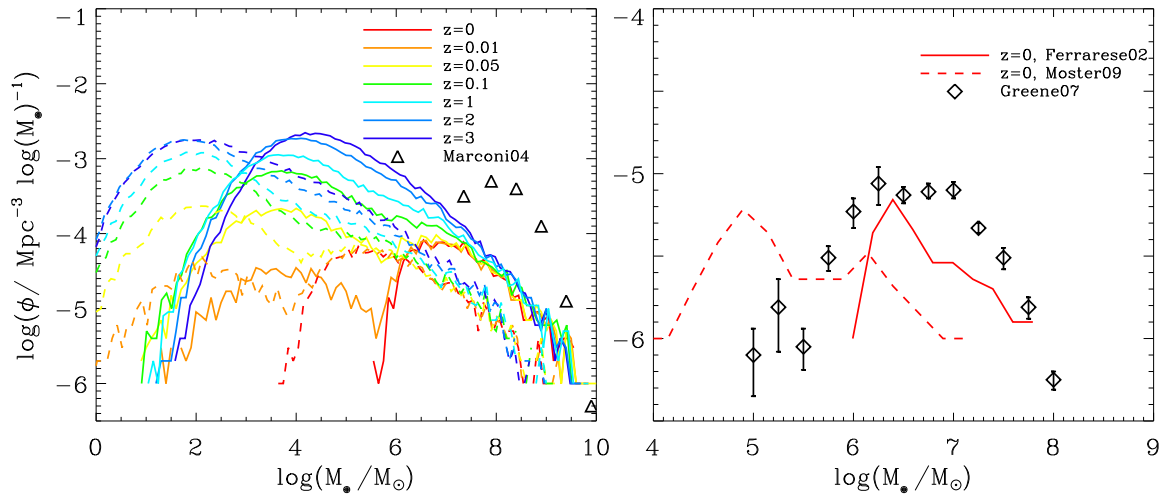, width=0.9\textwidth}
  \caption{Left: Evolution of the black hole mass function through $\Lambda$CDM-merging using 
using $M_{\bullet}$-$M_{DM}$-relation (solid lines) and using the galaxy population model (dashed lines). 
The triangles correspond to the local black hole mass function according to \citet{Marconi04}. Right: 
Same as in the left panel, but only for $z=0$. However, considered are only objects, that had undergone a merger 
event within the last $100 \cdot 10^6\ \mathrm{yrs}$ with a merger ratio smaller than $1:10$. The diamonds show 
the observed, present-day black hole mass function for active galaxies (\citealp{Greene07}).}
 {\label{Massfct_evol}}
\end{center}
\end{figure*}

In this section we investigate the evolution of the black hole mass function in our merger-driven model to test, if merging only provides an adequate description for black hole growth. 
Fig. \ref{Massfct_evol} shows the black hole mass function for different redshifts (different colors) assuming an 
initial log-normal scatter in black hole mass at $z=3$ and a growth of black holes via mergers only. 
The solid lines indicate the evolution of the black hole mass function assuming seeding according to 
the $M_{\bullet}$-$M_{\mathrm{DM}}$-relation (\citealp{Ferr02}) and the dashed lines show the black hole mass function 
based on the galaxy population model of \citet{Moster09a}. 
In the left panel we show the local observed black hole mass function (black triangles) derived from correlations 
between black hole mass and bulge luminosity or stellar velocity dispersion (\citealp{Marconi04}). This is in 
disagreement with the $z=0$ model prediction (red line), since we underestimate black hole masses. This implies that growth by gas accretion might be an important 
contribution to the overall formation process of black holes and should not be neglected. Furthermore, if we seed 
the black holes only at $z=1$, the disagreement with observations is - as expected - less pronounced than for $z=3$-seeding. 
But again, the black holes are not massive enough to reproduce the observations. This shows that even from $z=1$ till $z=0$  
merging only seems to be an unsufficient description for black hole growth.  
However, we want to point out most importantly that in this study we are \textit{not} trying to fit the present-day 
black hole mass function. In fact we want to show 
how the scatter evolution in black hole mass would be affected by merger events only. The possible influence of additional 
gas physics will be discussed in the next section.

The right panel of Fig. \ref{Massfct_evol} shows the black hole mass function for the 
two different seeding mechanisms only at $z=0$. However, here we have considered only galaxies that have undergone a 
merger with a merger ratio smaller than $1:10$ in the last $100 \cdot 10^6\ \mathrm{yrs}$. Therefore, they can be assumed to be in the active phase during this time. 
Our results deviate from observed values of the present-day black hole mass function for a sample of active galaxies ($8500$ objects from SDSS DR4, \citealp{Greene07}). On the one hand, this could be a consequence of underpredicting the overall black hole mass function as shown by the left panel, assuming the correct number of merger events. Alternatively, it might imply that merger events are not frequent enough to give an explanation for observed active galaxies and accretion is needed even without any merger event. 
However, if we assume, that also lower mass ratios ($< 1:100$) can trigger the activity of galaxies, we obtain a good agreement with observations. Also raising the time for one duty cycle would lead to a better consistence with observations. This could be justified by the fact, that within our assumed time for one duty cycle there could exist more active objects if we would consider \textit{galaxy} mergers, which happen later in time than mergers of isolated dark matter halos (which are taken into account in this study).

%*****************************************************************************************************
%*****************************************************************************************************
\section{Conclusions and Discussion}\label{conclusion}
%*****************************************************************************************************
%*****************************************************************************************************

In this paper we investigated the evolution of the intrinsic scatter in black hole mass under the assumption that the black holes 
only grow by mergers with other black holes during galaxy mergers. For many different idealized random merging models 
with (replenishment) and without (depletion) refilling from an external galaxy pool we find the following general results: 

\begin{enumerate}

\item The evolution of the black hole distribution can be well described within the framework of the central limit theorem (CLT). 
Independent 
of the initial distributions, e.g. log-normal or Schechter, of black holes the distribution always evolves into a normal 
distribution after a few merger generations. We consider all mergers independent of the mass ratio that galaxies $> 10^{4.7}$ M$_{\odot}$ experience.

\item \label{scatterdecrease} For all random merging models we found a decreasing scatter $\sigma$ with increasing 
merging generation $n$ and with increasing merger number $m$. As a consequence of the mass built-up during the merger 
events the scatter also decreases with increasing black hole mass. Motivated by the CLT, we can approximate the scatter 
dependence on the mean number of mergers by
\begin{equation}
\sigma_{\mathrm{merg}}(m) \approx \sigma_{\mathrm{ini}} \cdot (m+1)^{-a/2}.
\end{equation}
Here the exponent $a$ is a measure of the strength of the scatter decerease. For the different random merging models 
we find $0.30 < a < 0.61$ for a large number of mergers ($m >100$) independent of the initial scatter applied to black 
hole and bulge masses. In general, replenishment models show a weaker scatter decrease.

\item Considering either only major or only minor mergers for galaxy growth we found that minor mergers lead to a much 
stronger scatter decrease than major mergers; hence the smaller the mass ratio of merger events, the more rapidly the 
scatter decreases. This is in contrast to findings of \citet{Pengrandom}.

\item For different replenishment models we found that the higher the refill-ratio and the smaller the typical 
mass of black holes in the refill-pool, the more slowly the scatter decreases.

\end{enumerate}

Studying the effect of merging according to current structure formation models in $\Lambda$CDM-simulations, we 
find a qualitatively similar behaviour. The scatter decreases with the number of mergers ($a = 0.3$), and as a consequence 
it also decreases with cosmic time. This is also quantitatively consistent with the best-matching random merging model 
(refill-ratio $1/3$ and low mass refill-pool) at least for the limit of the low merger number range ($a = 0.34$), 
since in $\Lambda$CDM-merging the most massive galaxies experience only $\sim 50-60$ merger events. Therefore, 
the scatter evolution in $\Lambda$CDM-merging can be well approximated by a simple model assuming random merging of galaxies.

From the above results we can draw some implications about recent observations of high redshift black holes:
\begin{itemize}
\item For a simple merger driven growth of black holes we predict that the scatter in black hole mass must have been 
larger at higher redshift. Assuming an initial scatter of $0.6$ at $z=3$ the over-massive black holes investigated by 
\citet{Schramm}, \citet{Pengobs} and \citet{McLure06} would be within the $2-\sigma$ range of this large initial 
scatter. If these objects had on average $50-60$ dry merger events then we also obtain the present-day scatter value 
of $\sim 0.31$ for massive ellipticals \citep{Gueltekin09}. This shows that the observations of over-massive black 
holes at high redshifts are consistent with a population of galaxies that has a large scatter in black hole mass ($\sim 0.6$) 
at high redshifts being reduced through subsequent merging.

\item A further important advantage of our model is that we are not only able to explain over-massive black holes at high redshifts, but we can also account for the observed under-massive black holes at $z=2$ (\citealp{Alexander08, Shapiro09}). These objects would be  lying within the $2-\sigma$ range of the scatter at $z=2$ ($\sigma \approx 0.5$) assuming an initial scatter of $0.6$ at $z=3$. However, we want to point out, that models assuming an evolution of the $M_{\bullet}-M_{bulge}$ relation (from higher towards lower black hole masses with decreasing redshift) can be an explanation for over-massive black holes at high redshifts, but \textit{not} for the observed under-massive ones. 

\item We can explain the scatter evolution of model predictions from GALFORM (\citealp{Malbon}): 
the decreasing scatter with increasing black hole mass (see Fig. \ref{Comparison}). This is the 
consequence of the merger driven growth of black holes. More massive objects have undergone more 
merger events than less massive ones; this higher merger number leads to a stonger decrease in scatter. 
However, observations show a larger scatter in black hole masses, especially at the high mass end (\citealp{Tremaine02}, \citealp{Gueltekin09}). 
That might indicate that in the model of \citet{Malbon} either the number of merger events for massive objects 
was overestimated (unlikely) or that they started with a too small scatter in black hole mass at high redshift. 

\item Our finding of a decreasing scatter with increasing black hole mass as a consequence of subsequent merging are 
also consistent with the results of \citet{Gueltekin09}, who distinguish between two subsamples of ellipticals and non-ellipticals. 
For ellipticals they find a smaller scatter in black hole masses than for non-ellipticals. Referring elliptical 
galaxies mainly to the high-mass end and non-elliptical ones mainly to low-mass end, these observations are also 
consistent with our results. Furthermore, ellipticals presumably had more mergers. 

\end{itemize}

The aim of this paper is to investigate the scatter evolution of supermassive black hole for dry merger driven growth. Despite the 
interesting insights we have neglected another important physical growth mechanism, e.g. the accretion of gas onto black holes. 
The importance of an additional growth mechanism is also confirmed by the comparison of the black hole mass function resulting 
from merging only with observations at $z=0$: we under-estimate the masses of black holes. Furthermore, if we consider only active 
galaxies that have undergone a merger event in the last $100 \cdot 10^6\ \mathrm{yrs}$ with a merger ratio smaller than $1:10$, we again 
deviate from observations, which could be a consequence of underpredicting the over-all black hole mass function. Alternatively, it might 
indicate that merger events are not frequent enough to be an explanation for the observed active galaxies. This implies that there has to be additional gas accretion even if no merger event happened. 

Furthermore, according to the Soltan argument (\citealp{Soltan}) accretion on to massive black holes is the dominant source of energy 
produced by quasars ($L \sim \dot{M_{\bullet}} c^2$), which presumably indicates that the mass in black holes is mainly generated by 
accretion of gas. This gas accretion might be triggered by galaxy mergers and their ability to drive large amounts of gas to 
the centers of the galaxies and possibly feed a black hole (\citealp{Springel05, DiMatteo05, Johansson09, Hopkins08a}). It was shown in hydrodynamical galaxy-galaxy merger simulations with self-regulated black hole
 growth that after the active phase of the quasar, star formation and black hole growth are quenched by gas heating  
through energy release by the active quasar. Following the merger, the quasar host becomes a red and dead, massive elliptical galaxy 
without any further black hole growth through gas accretion (\citealp{Johansson09}). For these galaxies the further evolution 
of their black holes is mainly dominated by dry merger events.
However, gas-rich mergers and cold accreation flows seem to be the dominant growth mode for massive high redshift 
black holes, as the galaxies are more gas rich (\citealp{Khochfar06, Khochfar09}). The quasar activity peaks 
at $z\approx 2-3$ (e.g. \citealp{Hasinger06, Ueda05}). Therefore, this phase is most likely to be responsible for 
the scatter in black holes at this redshift (corresponding to the initial scatter in our models). According to \citet{Hasinger06} 
the emissivity of the high luminosity objects ($\log({L/L_{\odot}}) \ge 45$) drops by $\sim 10^2\ \mathrm{erg s}^{-1} \mathrm{Mpc}^{-3}$ between $z \sim 2$ and $z \sim 0$. Therefore, the subsequent evolution, at least for these massive black holes, might be driven mainly
by dry mergers as investigated in this paper. 
In addition, there is theoretical evidence that for $z \le 2$ gas is more efficiently heated by accretion shocks associated with 
gravitational heating in massive ($M_{\mathrm{DM}} \ge 10^{12} M_{\odot}$) galaxy halos (\citealp{Khochfar08, Dekel08, Naab09_2, Birnboim07}), suppressing further gas cooling and star formation as well as accretion onto the central supermassive black hole. 
Thus, 'red and dead' massive spheroids evolve starting at $z \sim 1$, where again merging may dominate the further growth of these black holes (\citealp{Dekel06, Khochfar09}).  
From disk galaxy merger simulations in \citet{Johansson09_2}, we know that overmassive black holes lying above the black 
hole-bulge relation do not evolve onto the relation considering only gas accretion driven growth. A solution for this 
problem could be that the evolution of overmassive black holes onto the relation might be caused mainly by growth through 
merger events leading to a scatter decrease in black hole mass as shown in this paper. That would also indicate, that gas 
accretion does not play the most important role in the evolution process of overmassive black holes, at least at late stages, 
where dry merging is the dominating process. 
 
Intentionally, in this paper we kept the physics very simple and focused only on the growth through merging in order to understand the influence of {\textit{merging}} on the evolution of the scatter in black hole mass. However, gas accretion is - even for $z<2$ - an important growth channel.
 The impact of gas accretion on the evolution of the scatter in the $M_{\bullet}-M_{bulge}$ relation is complicated and so far not easy to assess. To get an idea for such an influence, we refer the reader to \citet{Johansson09_2}. Using merger simulations of disk galaxies with gas ($3\cdot 10^9 M_{\odot} < M_{\mathrm{bulge}} < 10^{10} M_{\odot}$) they show two possible scenarios: Starting with galaxies on the relation leads to an increase of the scatter which gets even larger for higher gas fractions. However, considering initially galaxies below and above the relation causes a decrease of the scatter. The decrease seems to get stronger for lower gas fractions. In order to really understand the effect of gas physics we definitely need further investigations of this process in a statistical sense.

\section*{Acknowledgments}
This research was supported by the DFG Cluster of Excellence 'Origin and structure of the universe'. We would like to thank Ludwig Oser for providing the numerical simulations used in this study. In addition we thank the referee for his valuable comments and suggestions to strengthen the presentation of our paper.

\bibliographystyle{mn2e}
\bibliography{Literaturdatenbank}

\label{lastpage}

\end{document}